\documentclass[9pt,journal,twocolumn]{IEEEtran}

\usepackage[utf8]{inputenc}
\usepackage{graphicx}
\usepackage{epstopdf}
\usepackage{amsfonts}
\usepackage[cmex10]{amsmath}
\usepackage{mathtools}
\usepackage{framed,xcolor}
\usepackage{cite}
\usepackage{subfigure}
\usepackage{bm}
\definecolor{shadecolor}{gray}{0.90}
 
\DeclareMathAlphabet{\mathantt}{OT1}{antt}{li}{it}
\DeclareMathAlphabet{\mathantt}{OT1}{pzc}{m}{it}

\graphicspath{{./}}

\begin{document}
\title{Far-Field Inversion for the Deep Interior Scanning CubeSat}
\author{M.~Takala, P.~Bambach, J.~Deller, E.~Vilenius, M.~Wittig, H.~Lentz, H.~M.~Braun, M.~Kaasalainen, S.~Pursiainen
\thanks{M.~Takala (corresponding author) , M.~Kaasalainen and S.~Pursiainen are  with the Laboratory of Mathematics, Tampere University of Technology, P.O.\ Box 692, 33101 Tampere, Finland.}
\thanks{M.~Takala is with Laboratory of Pervasive Computing, Tampere University of Technology, P.O.\ Box 553, 33101 Tampere, Finland.}  
\thanks{P.~Bambach, J.~Deller and E.~Vilenius  are with Max Planck Institute for Solar System Research, Justus-von-Liebig-Weg 3, 37077 G\"ottingen, Germany.}
\thanks{M.~Wittig is with MEW-Aerospace, Hameln, Germany.} 
\thanks{H.~Lentz and H.~M.~Braun are with RST Radar Systemtechnik AG, Ebenaustrasse 8, 9413 Oberegg, Switzerland.} 
\thanks{Contact sampsa.pursiainen@tut.fi for further questions about this work.} } 
\maketitle
\begin{abstract}

This study aims at advancing mathematical and computational techniques for  reconstructing the  interior structure of a small Solar System body via Computed Radar Tomography (CRT). We introduce a far-field model for full-wave CRT and validate it numerically for an orbiting distance of 5 km using a synthetic 3D target asteroid  and sparse limited-angle data. As a potential future application of the proposed method, we consider the Deep Interior Scanning CUbeSat (DISCUS) concept in which the goal is to localize macroporosities inside a rubble pile near-Earth asteroid with two small spacecraft carrying a bistatic radar. 
\end{abstract}

\IEEEpeerreviewmaketitle 

\section{Introduction}
 
The aim of this study is to advance Computed Radar Tomography (CRT)  \cite{persico2014, devaney2012} for reconstructing the deep interior structure  of a small solar system body (SSSB) \cite{pursiainen2016, su2016, herique2016, kofman2015, asphaug2003}. The first such attempt, the Comet Nucleus Sounding Experiment by Radio-wave Transmission (CONSERT), was made as a part of the ESA's {\em Rosetta} mission to comet 67P/Churyumov-Gerasimenko. In CONSERT, a tomographic radar signal was transmitted between the orbiter and  the {\em Philae} lander \cite{kofman2007, kofman2015}.  At the moment,  several space organizations aim to rendezvous SSSBs. In 2018, the Osiris-REx by NASA  \cite{berry2013, lauretta2012}  and the Hayabusa-2   \cite{kawaguchi2008, tsuda2013} by Japan Aerospace Exploration Agency (JAXA) will arrive at  the asteroids 101955~Bennu and 162173~Ryugu (1999~JU3), respectively. Future CRT experiments have recently been planned for  the ESA's proposed Asteroid Impact Mission (AIM) to asteroid 65803 Didymos (1996~GT) \cite{herique2016, michel2016}.  
This paper introduces and validates a three-dimensional far-field extension for the full-wave CRT model presented in  \cite{pursiainen2016}. We use realistic parameter values, orbiting distance and target scaling in order to support the design of the future planetary missions. In spaceborne CRT, the ability to simulate and invert far-field data is essential, since inserting   a spacecraft into a stable SSSB orbit is a major challenge due to the low escape velocity of the SSSB. As the potential future application of the present inversion approach we consider the Deep Interior Scanning CUbesat (DISCUS) concept \cite{deller2017,bambach2017} in which two small spacecraft carrying a bistatic (dual-antenna) radar \cite{willis2008} record penetrating radar data at about  a few kilometers distance to a 260--600 m (Itokawa-size \cite{abe_mass_2006}) rubble pile near-Earth asteroid (NEA).  Using two CubeSats for a bistatic radar measurement  was first proposed for the AIM mission \cite{herique2016}.

Rubble-pile asteroids are celestial bodies composed of aggregates bound together by  gravitation and weak cohesion \cite{bottke2002,michel2015,richardson2005,polishook2016,sanchez2014,deller2015,deller2016}. Therefore, they are likely to contain macroporosity, e.g., internal voids or cracks. From the observed rotation period cut-off limit of about $>$$2.2\,\text{hours}$ for asteroids with diameters greater than 200--300~m, it has been concluded that these bodies are not monolithic, but in the vast majority rubble pile asteroids \cite{Pravec2000}. The estimated density for the rubble piles suggests that macroporosity exists \cite{carry_density_2012} but the actual proof is still missing. The radar onboard DISCUS would use a \mbox{20--50~MHz} center frequency which is advantageous regarding both void detection \cite{binzel2005, daniels2004} and also the measurement noise due to the Sun  \cite{burke2009,stone2000,kraus1967}. As an independent mission, DISCUS would demonstrate a new and affordable mission concept to gain knowledge of the inside of NEAs. 

In the numerical experiments, we investigated a three-dimensional asteroid model containing deep interior anomalies and a surface dust layer.  The inversion accuracy and reliability were explored for several different noise levels and also for sparse limited-angle data, i.e., a situation in which the trajectory of the spacecraft does not allow measurements from all directions.  Full-wave CRT was applied in order to maximize the imaging quality and to allow the sparsity of the measurements which is vital for the many limitations of the space missions, such as low data transfer rates and comparably short instrument lifetime \cite{agrawal2014, doody2010}. Volumetric full-wave CRT in a realistic geometry is a computationally challenging imaging technique which requires 3D wave simulation and inversion of a large system of equations. In order to achieve a sufficiently short computation time, a state-of-the-art cluster of graphics processing units (GPUs) was applied in the forward simulation. The waveform data were linearized and inverted using a high-end dual-processor workstation computer. 

The results obtained suggest that the proposed mathematical model can be applied to invert full-wave far-field data for realistic asteroid sizes and shapes. Furthermore, based on the results, it seems that the existing background noise in the solar system and the expected level of modeling errors allow detection of macroporosities from the planned orbiting distance. The bistatic and stepped-frequency measurement techniques  present in the DISCUS concept were found to be vital for improving the signal-to-noise ratio and for reducing the effects of the measurement noise and modeling errors. A comparison between full- and limited-angle CRT results suggests that the interior structure of the  targeted NEA with a typical spin orientation can be reconstructed without requiring an orbital plane that includes shadow phases of the spacecraft and without the need to  cover the entire surface of the SSSB. 

This paper is organized as follows. Section \ref{mm} briefly describes the DISCUS radar,  numerical experiments and the full-wave CRT model for the far-field measurements. Sections \ref{r} and \ref{d}  include the results and the discussion. 

\section{Materials and Methods}
\label{mm}

\subsection{DISCUS radar}

The DISCUS concept comprises two identical CubeSats. Both carry an identical radar instrument capable of a 10 W signal transmission.  The CubeSats are equipped with a half-wavelength dipole antenna with a center frequency $f_c$ between 20 and 50 MHz and a bandwidth $B$ of at least 2 MHz. The bistatic measurement approach is used, since 
the scattering waves can be headed away from the transmitter spacecraft (Figure \ref{bistatic_measurement}).  That is, one of the CubeSats both transmits and receives the signal, and the other one serves as an additional receiver. 

The stepped-frequency measurement technique is applied  \cite{iizuka1984,gill2001,paulose1994}. That is, the signal is a pulse sequence of narrow frequency lines  $\psi_1, \psi_2, \ldots, \psi_N$ which allow one to approximate a given function $f$ via the sum \begin{equation} f = \sum_{\ell = 1}^{N} c_\ell \psi_\ell. \label{f_formula} \end{equation} If $\varphi_\ell$ is the received signal  corresponding to frequency line $\psi_\ell$, then the data $g$ resulting from a transmission $f$ is approximately given by $g = \sum_{\ell = 1}^{N} c_\ell \varphi_\ell$ with $c_1, c_2, \ldots, c_N$ following from (\ref{f_formula}). Consequently, the data $g$ for any transmission $f$ within the given frequency range can be approximated, if the function pairs $\psi_\ell$ and $\varphi_\ell$ for $\ell = 1, 2, \ldots, N$ are given, i.e., if the stepped-frequency measurement data are available. Recording the frequency lines separately is advantageous with regard to the measurement accuracy, since the signal-to-noise ratio between the received power and the noise power is inversely proportional to the bandwidth of a single line.

The CubeSats follow their target asteroid in a polygonal plane which will be nearly perpendicular to the ecliptic of the solar system. The signals will be transmitted and captured via the two half-wavelength dipole antennas which will be pointed towards the Sun during the measurement to suppress the effect of the solar radiation (Figure \ref{bistatic_measurement}). In order to optimize the radar performance for the target NEA's interior structure, the spacecraft will be equipped with a camera so that an accurate optical surface model can be created. 

\subsection{Numerical Modeling}

In this study, we validate the radar concept of DISCUS numerically using the publicly available surface model of the asteroid 1998 KY26 scaled to 550 m diameter.  The center frequency and bandwidth of the radar measurement are assumed to be $f_c = 20$ MHz (wavelength $\lambda = 15$ m in vacuum) and $B \approx 2.4$ MHz, respectively. The essential  model parameters and their values can be found in Table \ref{scaling}. In the numerical simulation, we use a scalable unitless parameter presentation. The unitless values can be scaled to SI-units as shown in Table \ref{scaling_2}.

The signal power $P_{RX}$ (dB) received by the radar is estimated via the equation 
\begin{equation}
P_{RX} = P_{TX} + G_{TX} + S + G_{RX},    
\end{equation}
where $P_{TX} = 0$ dB is the power transmitted, $G_{TX} = G_{RX} = 2.15$ dBi (1.64) follows from the gain of the half-wavelength dipole antenna and $S$ is the signal power at the receiver location obtained through an isotropic radiator model, i.e., with an isotropic source and effective antenna aperture $A_{eff} = \lambda^2/(4 \pi)$.

In the stepped frequency measurement, the bandwidth of a single frequency line is given approximately by $B_\ell = 1 / T_\ell$, where $T_\ell$ is the pulse duration. We assume that the data is collected at a 5 km distance to the target asteroid. The duration of a single line is set to be $T_\ell = 32$ $\mu$s, i.e., 96 \% of the signal travel-time, in order to minimize the corresponding bandwidth (here $B_\ell = 31$ kHz) and, thereby, also the relative amplitude of the measurement noise. 

\begin{table*} \begin{center}
\caption{ The present numerical model in unitless format and  SI-units. The spatial parameters are also given relative to center wavelength in vacuum ($\lambda = 15$ m). The conductivity value and loss rate of porous basalt can be found in \cite{olhoeft1981,kofman2012}. Here $B$ is an estimate for the radar bandwidth which allows transmitting the signal pulse with errors below -20 dB. Center frequency $f_c$ is used in the calculation of the antenna aperture.  \label{scaling}}

\begin{tabular}{llll}
Item & Unitless & SI-units & Relative to $\lambda$\\
\hline 
Asteroid diameter & 0.262 &  550 m  & 36.7 \\
Geometry scaling factor $s$ & 1 & 2100 m & 140 \\
Conductivity $\sigma$ & 24 & 3E-5 S/m & \\
Attenuation rate & 52.5 dB/(unit length) &  25 dB/km \\
Orbiting distance & 2.4 & 5 km  & 330 \\
Void diameter  & 0.029--0.048 &  60--100 m  & 4--6.7\\
Dust layer thickness & 0.019  & 40 m  & 2.7 \\ 
Void $\varepsilon_r$ & 1 & 1 \\
Dust $\varepsilon_r$ & 3& 3 \\ 
Body $\varepsilon_r$ & 4 &  4 \\
Final pulse duration after processing $T_0$ &0.12  & 0.84 $\mu$s \\
Radar bandwidth $B \approx 2/T_0$  & $16.7$ 1/(unit time) &  $2.4$ MHz  \\
Center frequency $f_c$ & 140 &  20 MHz \\ 
Center wavelength in vacuum $\lambda$ & 7.1e-2 & 15 m & 1 \\
Final data duration after processing $T$ & 0.7 & 5.0 $\mu$s  \\
Frequency line duration $T_\ell$ & 4.6E-3& 32 $\mu$s  \\
Frequency line bandwidth $B_\ell$  & 217 1/(unit time) & 31 kHz \\
 \hline
\end{tabular} \end{center} 
\end{table*}

\begin{table} \begin{center}
\caption{Formulas for scaling between the unitless and SI-unit expressions. In these, the permittivity and magnetic permeability of the vacuum are given by $\varepsilon_0 = 8.85 \cdot 10^{-12}$ F/m, and $\mu_0 = 4 \pi \cdot 10^{-7}$ b/m, respectively, and $s$ (meters) denotes the spatial scaling factor.\label{scaling_2}}

\begin{tabular}{lll}
Item & Unitless & SI-units    \\
\hline
Dielectric permittitivity & $\varepsilon_r$  & $\varepsilon_r$ \\  Electrical conductivity & $\sigma$ &  $(\mu_0 / \varepsilon_0)^{-1/2} s^{-1} \sigma$ \\  
Position   & $\vec{x}$ & $s {\vec x}$ \\  
Time & $t$ & $(\varepsilon_0  \mu_0)^{1/2} s t$ \\ Frequency & $t^{-1}$ &$(\varepsilon_0  \mu_0)^{-1/2} s^{-1} t^{-1}$ \\ Velocity (${\mathsf c} = \varepsilon_r$) & ${\mathsf c}$ & $(\varepsilon_0  \mu_0 )^{-1/2} {\mathsf c}$ \\
 \hline
\end{tabular} \end{center} 
\end{table}

\subsubsection{Permittivity and Conductivity}

The present inverse problem of the CRT is to reconstruct the target asteroid's internal dielectric permittivity distribution $\varepsilon_r$  which contains, in this study, a 40 m surface  layer (e.g.\ dust or sand) with $\varepsilon_r = 3$  and three 60--100 m interior voids with  $\varepsilon_r = 1$ (vacuum), respectively. Otherwise, $\varepsilon_r$ is assumed to be 4, which is typical, e.g., for Kaolinite and Dunite \cite{herique2002}. 

The motivation for using this synthetic and relatively simple permittivity distribution was to enable validating  the current far-field inversion approach both for the surface and deep interior structures with an exact measure for the inversion accuracy. The size of the details is mainly determined by the estimated tomography resolution. According to the recent studies, macroporosity structures of the present size and permittivity range can exist in rubble pile asteroids \cite{carry_density_2012,michel2015}.

The electrical conductivity distribution inside the asteroid was assumed to be an unknown nuisance parameter, i.e., not of primary interest.  It was set to be $\sigma = 5 \varepsilon_r$ matching roughly to 3E-5 S/m and a loss rate of around 25 dB/km  which is typical, e.g.,  for porous basalt (pyroxene) \cite{olhoeft1981,kofman2012}. Outside the asteroid the $\sigma$ was assumed to be zero.

\subsubsection{Measurements} 

\begin{figure}
\begin{center}
\begin{minipage}{3.5cm}
\begin{center}
\includegraphics[width=3cm]{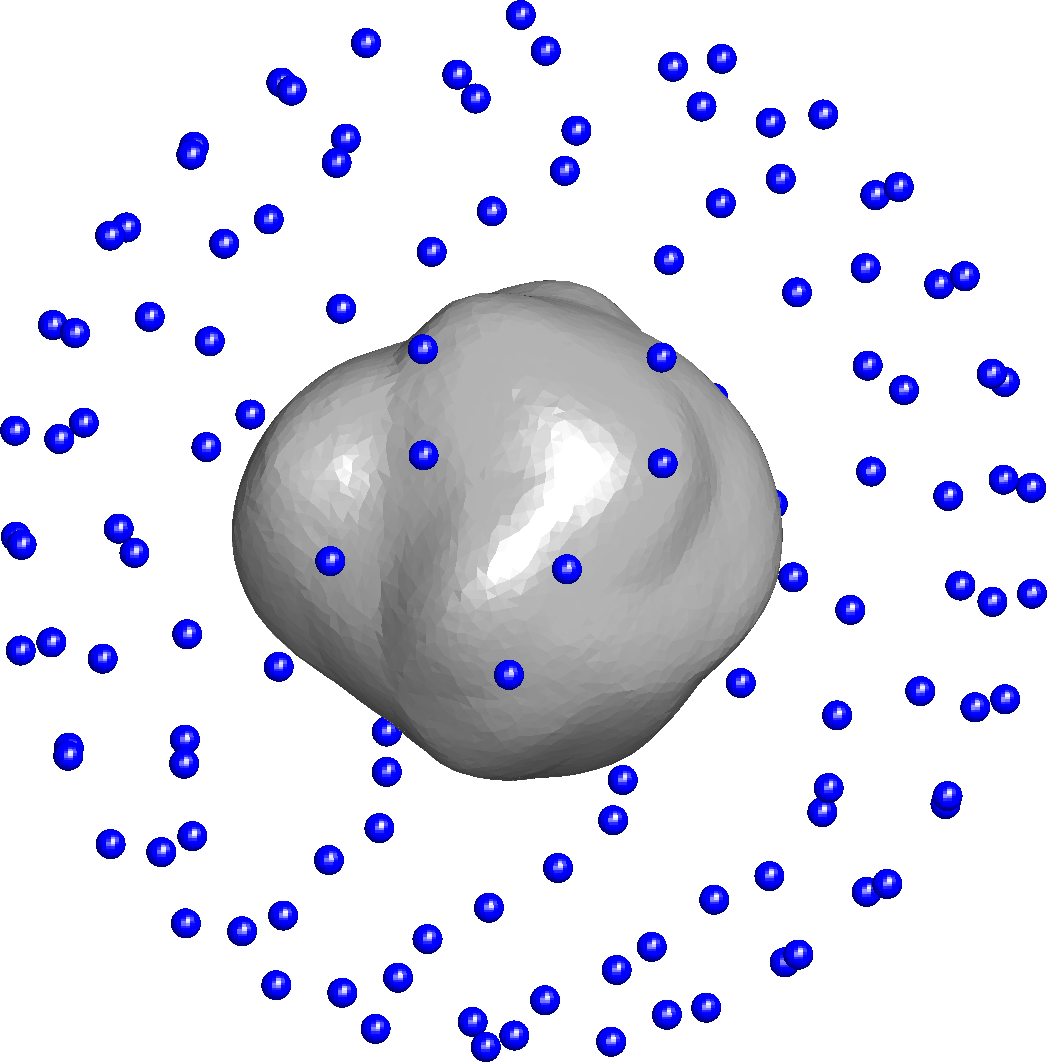} \\ (A) : $\omega = 90^\circ$ 
\end{center}
\end{minipage}
\begin{minipage}{3.5cm}
\begin{center}
\includegraphics[width=3cm]{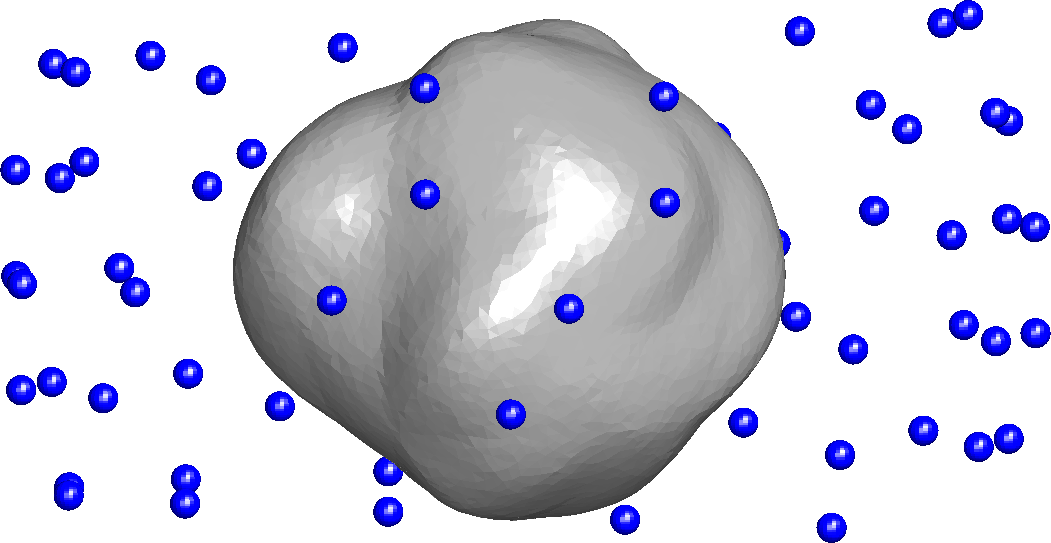} \\ (B) : $\omega = 30^\circ$ \\ \vskip0.2cm
\includegraphics[width=3cm]{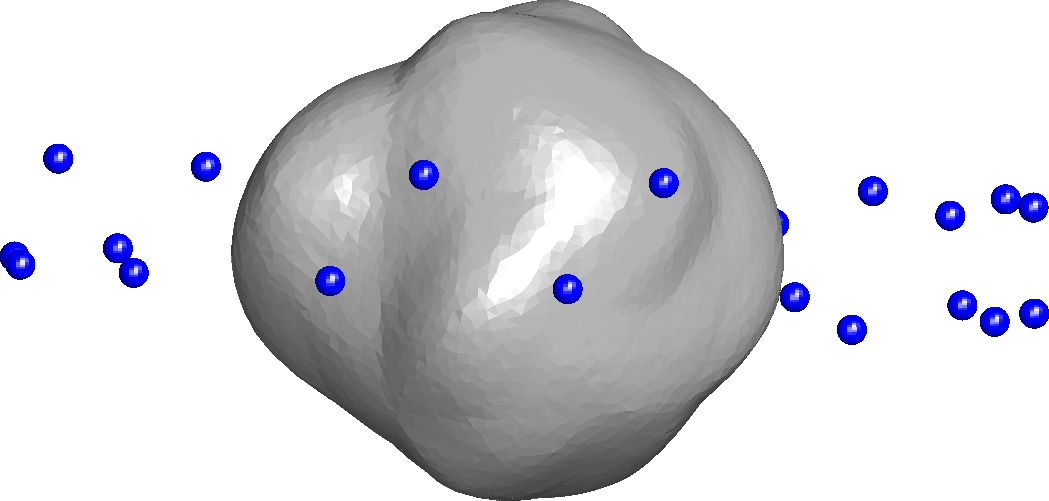} \\ (C) :  $\omega = 10^\circ$ 
\end{center}
\end{minipage}
\end{center}
\caption{
A schematic (non-scaled) illustration showing the sparse full-angle configuration (A) of 128  measurement points (left) and two sparse limited-angle configurations (B) and (C) of 64 and 24 points (right), respectively. The angle $\omega$ between the orbiting plane normal and the asteroid spin is  90, 30 and 15 degrees.  The limited-angle configurations include an aperture around the z-axis. \label{measurement_points}}
\end{figure}

The data are simulated for 128 measurement points (Figure \ref{measurement_points}) evenly distributed on an origin centric sphere. The angle $\omega$ between the asteroid spin and the normal of the orbiting plane determines the coverage of the measurements. We investigate the following three different cases (A) $\omega = 90^\circ$, (B)  $\omega = 30^\circ$ and (C)  $\omega = 10^\circ$. In (A), a full-angle dataset can be recorded, meaning that the spacecraft will form a dense network of points enclosing the whole asteroid. The limited-angle cases (B) and (C), include an aperture of $90^\circ - \omega$ around the spin axis. 

The following two measurement approaches were compared: 
\begin{enumerate}
\item {\em Monostatic measurement}. A single spacecraft both transmits and receives the signal. The measurement is made each time at the point of the transmission. The total number of measurement positions is 128, 64, and 24 for (A), (B) and (C), respectively.
\item {\em Bistatic measurement}. Two spacecraft are used; one transmits the signal and both record the backscattered wave (Figure \ref{measurement_points}). Each measurement is made simultaneously at two different positions. The additional measurement is made at the point closest to 25 degrees apart from the transmission location with respect to the asteroid's center of mass.  The total number of transmission/measurement points is 128/256, 64/128 and 24/48 for (A), (B) and (C), respectively. In each case, the total number of unique measurement positions in the point set is the same as in the respective monostatic case. 
\end{enumerate}

\begin{figure}
\begin{center}
\begin{minipage}{5cm}
\includegraphics[height=3.5cm]{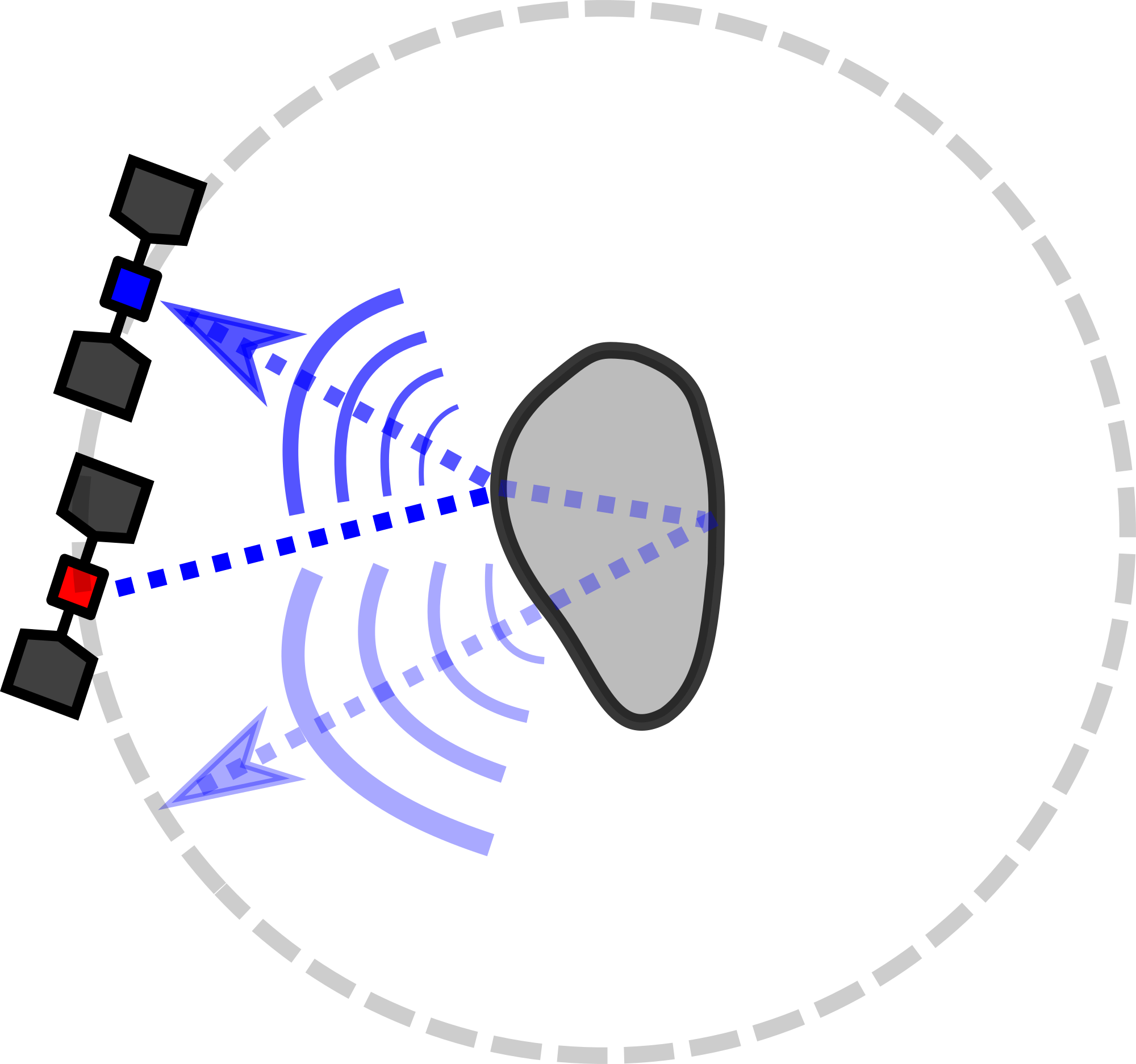}
\end{minipage}
\begin{minipage}{2cm}
\includegraphics[height=3.7cm]{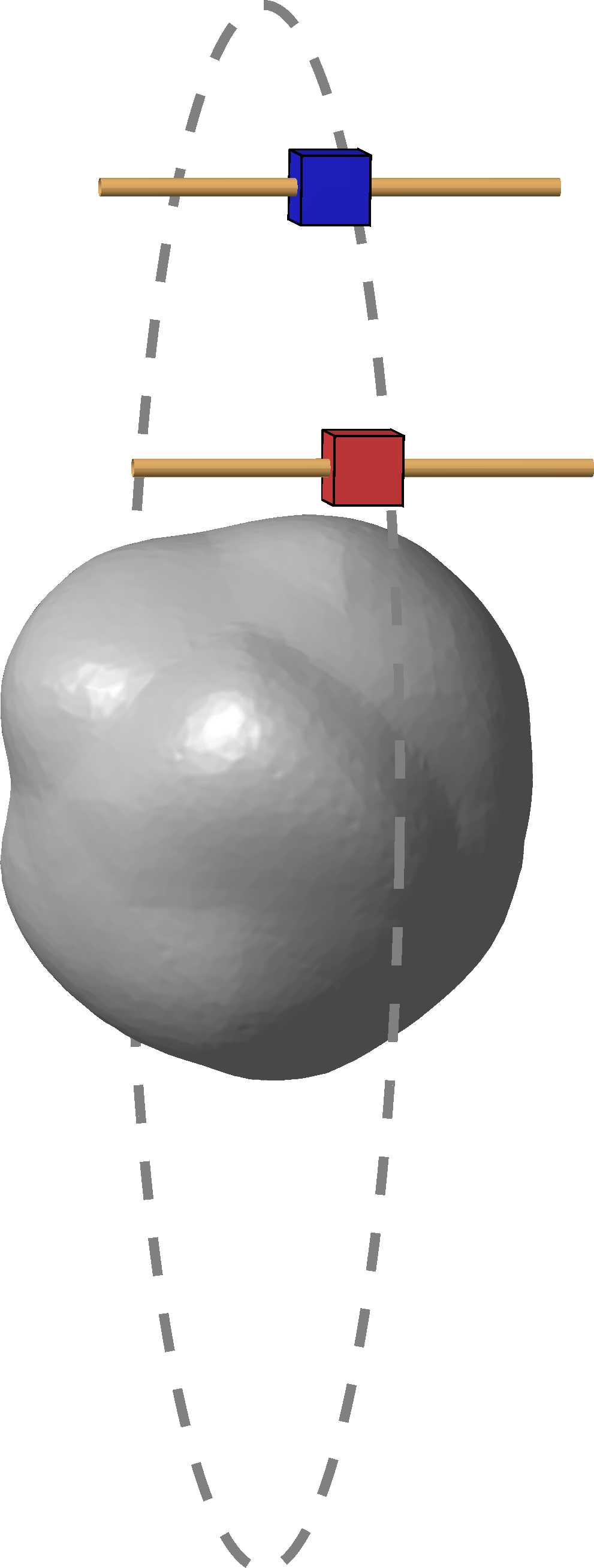}
\end{minipage}
\end{center}
\caption{{\bf Left:} A bistatic measurement approach is used, since the scattering waves can be headed away from the transmitter spacecraft. One spacecraft (red) is utilized to both transmit and receiver the signal. The other one (blue) serves as an additional receiver. {\bf Right:} The spacecraft are assumed to orbit their target asteroid in a plane which will be nearly perpendicular to the ecliptic of the solar system. Each spacecraft is equipped with a single half wavelength dipole antenna (yellow) which will be pointed towards the Sun during the measurement in order to suppress the effect of the solar radiation. That is, the Sun is assumed to be located in the direction of the normal of the orbiting plane. 
 \label{bistatic_measurement} }
\end{figure}

\subsection{Far-Field Forward Model for CRT}

The forward problem of CRT is to predict the voltage of the antenna given the computation geometry and the unknown parameters. In this study, we model a sparse set of (simulated) measurements in the spatio-temporal domain $[0, T] \times \Omega$ applying the  wave propagation model presented in Appendix. 

\subsubsection{Weak form}  

The signal transmission at point $\vec{p}$ is modeled as a point source of the form  \begin{equation}\label{point_source} \frac{\partial  \vec{f}}{\partial t}(t, \vec{x})\,  = \frac{\partial \vec{h}}{\partial t}(t)  \, \delta(\vec{x} - \vec{p})   \quad \hbox{with} \quad   h(0) = \frac{\partial \vec{h}}{\partial t}(0) = 0, \end{equation} where  $\vec{h}$ denotes the dependence of $\vec{f}$ on time and $\delta(\vec{x} - \vec{p})$ is the Dirac's delta function satisfying $\int q(\vec{x})\delta (\vec{x} - \vec{p}) \, d V = q(\vec{p})$ for any sufficiently regular function $q$. Physically $\vec{f}$ can be interpreted as the current density of the antenna (Appendix).  

The $i$-th component $E = E_i$ electric field $\vec{E} = E_i$ evoked by $\vec{f}$ satisfies the following weak form:
{\setlength\arraycolsep{2pt} \begin{eqnarray} \label{weak_form_1}  a ( E, {\bf g} ; {\bf w}; \Omega)  & = &    0, \label{weak_form_2a}\\
 b ( E, {\bf g}; v; \Omega)  & = &    - \langle  f, v ;\Omega \rangle  \label{weak_form_2}
\end{eqnarray}}
where ${\bf g} =  (\vec{g}^{\,(1)}, \vec{g}^{\, (2)}, \vec{g}^{\, (3)})$ with  $\vec{g^{\,(j)}} = \int_0^t \nabla E_i (\tau, \vec{x}) \, d \tau$ for $j = 1, 2, 3$ and $\langle  f, v ;\Omega \rangle = \int_\Omega f v \, \hbox{d} V$.  The bilinear forms $a$ and $b$ correspond  to the right-hand sides of the equations (\ref{wf1}) and (\ref{wf2}), respectively, and $v  \in H^1(\Omega)$, ${\bf w} = (\vec{w}^{(1)}, \vec{w}^{(2)}, \vec{w}^{(3})$ with $\vec{w}^{(i)} \in  [L_2(\Omega)]^3$  are test functions. Under regular enough initial conditions this weak form \cite{pursiainen2016} has a unique solution $E : [0,T] \to H^1(\Omega)$ \cite{evans1998}. 

\begin{figure}
\begin{center}
\begin{minipage}{4.2cm}
\includegraphics[width=3.9cm]{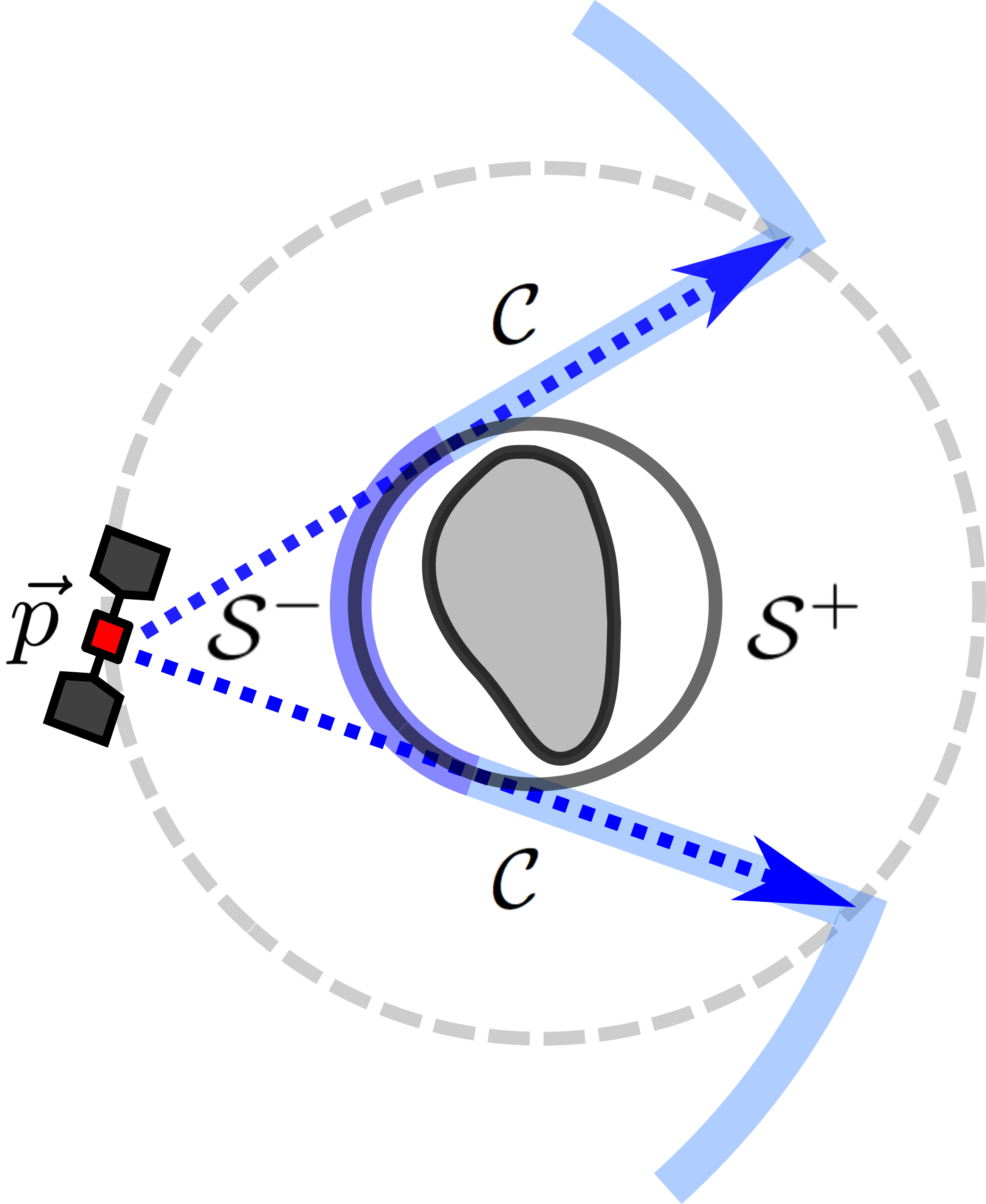}
\end{minipage}
\end{center}
\caption{ The incident field needs to be evaluated only on its first arrival at $\mathcal{S}$ (dark gray circle), that is, on the subset $\mathcal{S}^- = \{ \vec{x} \in \mathcal{S} \, | \, (\vec{x} - \vec{p}) \cdot \vec{n} < 0 \} $ (dark blue). On the remaining part  $\mathcal{S}^+ = \mathcal{S} \setminus \mathcal{S}^-$, it is set to be zero. Surface $\mathcal{C}$  (dark and light blue) results from extending the surface $\mathcal{S}^-$ with the tangent cone (light blue) of $\mathcal{S}$  which intersects $\vec{p}$, and further with  a $\vec{p}$-centric sphere with a radius larger than $T$. It holds that $\langle \vec{g}_{\mathtt{I}}, v ; \mathcal{C} \rangle = \langle \vec{g}_{\mathtt{I}}, v ; \mathcal{S}^- \rangle$. Namely, the integrand in (\ref{s_integral}) is zero on the tangent cone and within the distance $> T$ from $\vec{p}$ the incident wave is zero for $t \in [0, T]$. \label{far_field_calculation}}
\end{figure}

For modeling the far-field, we assume that the domain $\Omega$ consists of two sub-domains: an outer part $\Omega_1$ and an enclosed ball $\Omega_2$ containing the target asteroid. The spherical inner boundary is denoted with $\mathcal{S} = \Omega_1 \cap \Omega_2$ (Figure \ref{far_field_calculation}). The spacecraft position $\vec{p}$ is assumed to be located outside $\Omega$. 

\subsubsection{Incident and Scattered Field} 

The total field $E$ is expressed as the sum of the incident and scattered field, i.e., $E = E_{\mathtt I} + E_{\mathtt S}$, where the incident field $E_{\mathtt I}$ emanates from the source and vanishes in the interior of $\Omega_2$, that is, in $\Omega_{2} \setminus \mathcal{S}$. The scattered field $E_{\mathtt S}$ is the total field in $\Omega_{2}$, and its restriction to the surface  $\mathcal{S}$ is utilized in calculating the measured far-field. The incident field $E_{\mathtt I}$ satisfies the following weak form in $\Omega_{1}$:
{\setlength\arraycolsep{2pt} \begin{eqnarray}   \label{weak_form_3} a ( E_{\mathtt I}, {\bf g}_{\mathtt I};  {\bf w}; \Omega_1)  & = &    0, \\
 b ( E_{\mathtt I}, {\bf g}_{\mathtt I}; v; \Omega_1)  & = &    - \langle  f, v ; \Omega_1 \rangle + \langle  {\bf g}_{\mathtt I }, v ;\mathcal{S} \rangle , \label{weak_form_4}
\end{eqnarray}}
in which 
\begin{equation}
\label{surface_source}
 \langle  {\bf g}_{\mathtt{I}}, v  ; \mathcal{S} \rangle   =  \int_\mathtt{S} (\vec{g}_{\mathtt{I}} \cdot \vec{n}) v  \, \hbox{d} S
\end{equation}
with $\vec{g}_\mathtt{I} = \int_0^t \nabla E_\mathtt{I} (\tau, \vec{x}) \, d \tau$. The inner product $\langle  {\bf g}_{\mathtt{I}}, v  ; \mathcal{S} \rangle$ corresponds to the nonzero surface term in the integral 
\begin{equation} \int_{\Omega_i} (\nabla \cdot \vec{g}_{\mathtt{I}} ) \, v \, d V =  (-1)^{i} \int_\mathtt{S} (\vec{g}_{\mathtt{I}}\cdot \vec{n}) v  \, \hbox{d} S -   \int_{\Omega_i}  \vec{g}_{\mathtt{I}}\cdot \nabla v \, \hbox{d} V, \end{equation}
where $\vec{n}$ denotes the outward unit normal of $\mathcal{S}$ and $j = 1, 2$. For the scattered field $E_{\mathtt S}$, it holds $E_{\mathtt S} = E - E_{\mathtt I}$. Thus, its weak form in $\Omega_i$ for $i = 1, 2$ can be obtained by subtracting both sides of (\ref{weak_form_3}), (\ref{weak_form_4}) from  (\ref{weak_form_1}), (\ref{weak_form_2}) written for $\Omega_1$ and $\Omega_2$, respectively, taking into account that $\langle f, v; \Omega_2 \rangle_2 = 0$ and that the incident field vanishes in  $\Omega_{2} \setminus \mathcal{S}$. It follows that
{\setlength\arraycolsep{2pt} \begin{eqnarray}   \label{weak_form_5} a ( E_{\mathtt S}, {\bf g}_{\mathtt S};  {\bf w}; \Omega_i)  & = &    0, \\
 b ( E_{\mathtt S}, {\bf g}_{\mathtt S}; v; \Omega_i)  & = &    - \langle  {\bf g}_{\mathtt I}, v ;\mathcal{S} \rangle  \label{weak_form_6}
\end{eqnarray}}
for $i = 1 , 2$. Summing both sides of (\ref{weak_form_5}), (\ref{weak_form_6}) for $\Omega_1$ and $\Omega_2$ together leads to the following  full domain  weak form for the scattered field:
{\setlength\arraycolsep{2pt} \begin{eqnarray}   \label{weak_form_7} a ( E_{\mathtt S}, {\bf g}_{\mathtt S};  {\bf w}; \Omega)  & = &    0, \\
 b ( E_{\mathtt S}, {\bf g}_{\mathtt S}; v; \Omega)  & = &    - 2 \,  \langle  {\bf g}_{\mathtt I }, v ;\mathcal{S} \rangle .  \label{weak_form_8}
\end{eqnarray}}
This formulation is otherwise similar to (\ref{weak_form_1})--(\ref{weak_form_2}), but instead of a single point, the source function $\bf{g}_{\mathtt I}$ is evaluated on the sphere $\mathcal{S}$.  

\subsubsection{Incident Far-Field} In empty space ($\varepsilon_r = 1 $ and $\sigma = 0$), the incident field for a monopolar (isotropic) point source  (\ref{point_source}) placed at $\vec{p}$ can be expressed as the following convolution: 
\begin{equation}
E_{\mathtt I} = - \mathcal{G} \ast_\tau \frac{\partial  h}{\partial t}= - \frac{ 1   }{4 \pi |\vec{x} - \vec{p}|} \left[ \frac{\partial h}{\partial t}\right],
\label{incident_field}
\end{equation}
where $h \ast_\tau k(t, \vec{x}) = \int_0^\infty h (t - \tau, \vec{x}) k(\tau, \vec{x}) \, d \tau$, $[ {\partial h}/{\partial t}] $ is a retarded signal evaluated at $t -  |\vec{x} - \vec{p}|$ and $\mathcal{G} = {\delta(t - |\vec{x} - \vec{p}|)}/{(4 \pi |\vec{x} - \vec{p}|)}$ is a  Green's function with $\delta$ denoting  the Dirac's delta function defined with respect to time. That is,  $h(t, \vec{x}) = \int_0^\infty h(t-\tau, \vec{x}) \delta(\tau) \, d \tau$ for any sufficiently regular $h$.  It follows from (\ref{incident_field}) through a  straightforward calculation  that 
{\setlength\arraycolsep{1pt}  \begin{eqnarray}
\nabla E_{\mathtt I} & = &   \frac{(\vec{x} - \vec{p})}{4 \pi |\vec{x} - \vec{p}|^3} \left[ \frac{\partial h}{\partial t}\right]  + \frac{(\vec{x} - \vec{p}) }{4 \pi |\vec{x} - \vec{p}|^2} \left[ \frac{\partial^2 h}{\partial t^2}\right] \\
\vec{g}_{\mathtt I}  & = &    \int_0^t \! \nabla u_{\mathtt{I}} d \tau \! = \! \frac{\vec{x}  -  \vec{p} }{4 \pi |\vec{x} -  \vec{p}|^3} [{h}]  \! + \!  \frac{\vec{x}  - \vec{p}}{4 \pi |\vec{x}  -  \vec{p}|^2} \left[ \frac{\partial h}{\partial t}\right] \! .
\end{eqnarray}}

The incident field $E_{\mathtt I}$ needs to be evaluated only on its first arrival at $\mathcal{S}$, that is, in the subset $\mathcal{S}^- = \{ \vec{x} \in \mathcal{S} \, | \, (\vec{x} - \vec{p}) \cdot \vec{n} < 0 \} $. In the remaining part  $\mathcal{S}^+ = \mathcal{S} \setminus \mathcal{S}^-$, $E_{\mathtt I}$ is set to be zero. Hence, it follows that $\langle \vec{g}_{\mathtt{I}}, v ; \mathcal{S}^- \rangle = \langle \vec{g}_{\mathtt{I}}, v ; \mathcal{S} \rangle$ with 
\begin{equation}
\langle \vec{g}_{\mathtt{I}}, v ; \mathcal{S}^- \rangle \! =  \int_{\mathcal{S}^{+}} \! \left( \frac{( \vec{x}  \! - \! \vec{p} ) \cdot \vec{n}}{4 \pi |\vec{x} \!  - \!   \vec{p}|^3} [{h}]  \! + \!  \frac{(\vec{x}  \! - \!  \vec{p}) \cdot \vec{n}}{4 \pi |\vec{x}  \! - \!  \vec{p}|^2} \left[ \frac{\partial h}{\partial t}\right] \right)  v \,  \hbox{d} S .
\label{s_integral}
\end{equation}
This can be verified by extending the surface $\mathcal{S}^-$ with the tangent cone of $\mathcal{S}$  which intersects $\vec{p}$, and further with  a $\vec{p}$-centric sphere with a radius larger than $T$. For the resulting surface $\mathcal{C}$,  it holds that $\langle \vec{g}_{\mathtt{I}}, v ; \mathcal{C} \rangle = \langle \vec{g}_{\mathtt{I}}, v ; \mathcal{S}^- \rangle$. Namely, the integrand in (\ref{s_integral}) is zero on the tangent cone and within the distance $> T$ from $\vec{p}$ the incident wave is zero for $t \in [0, T]$. 

\subsubsection{Scattered Far-Field}

Following from the Green's integral identities \cite{jackson2007}, the empty space  ($\varepsilon_r = 1, \sigma = 0$) wave radiating out of $\mathcal{S}$  satisfies the following Kirchhoff integral equation, which we utilize to extrapolate the scattered far-field (total field) at $\vec{p}$: 
{\setlength\arraycolsep{2pt}  \begin{eqnarray}
\label{far_field_measurement_1}
E(\vec{p}) & = & \int _{\mathcal{S}}  [E_{\mathcal{S}}] \frac{\partial \mathcal{G}}{\partial \vec{n}} \, \hbox{d} S -  \int_\mathtt{S} \mathcal{G} \left[ \frac{\partial {E_{\mathcal{S}}}}{\partial \vec{n}}\right]    \hbox{d} S \\ \label{far_field_measurement_2} & = &   -  \frac{1}{4 \pi} \int_\mathtt{S} \frac{(\vec{x}  -   \vec{p}) \cdot \vec{n}}{ |\vec{x} -   \vec{p}|^2} \left(    \frac{[E_{\mathcal{S}}]}{ |\vec{x}  -   \vec{p}|}    +   \left[ \frac{\partial {E}_{\mathcal{S}}}{\partial t}\right]  \right)  \hbox{d} S \nonumber \\ & & -   \frac{1}{4 \pi} \int_\mathtt{S} \frac{1}{ |\vec{x}  -   \vec{p}|}   \left[ \frac{\partial {E_{\mathcal{S}}}}{\partial \vec{n}}\right]   \hbox{d} S. \label{far_field_measurement_3}
\end{eqnarray}}

\subsection{Linearized Forward Model, Signal and Noise}

We use the following  linearized forward model \cite{pursiainen2016} in which the point of linearization is a constant background permittivity $\varepsilon_r^{(\hbox{\scriptsize bg})} $: 
\begin{equation}
{\bf y}  = {\bf L} {\bf x} +  {\bf y}^{(\hbox{\scriptsize bg})} + {\bf n}.  
\end{equation}
The vectors ${\bf y}$  and  ${\bf y}^{(\hbox{\scriptsize bg})}$ contain the measured and simulated data for $\varepsilon_r$ and  $\varepsilon_r^{(\hbox{\scriptsize bg})}$, respectively, ${\bf x}$ is the coordinate vector for $\varepsilon_r$, ${\bf L}$ denotes the Jacobian matrix resulting from the linearization, and ${\bf n}$ contains both the measurement and forward modeling errors. 

A single Blackman-Harris window $h(t)$ \cite{harris1978,nuttall1981}  is used as the signal pulse. That is, 
{\setlength\arraycolsep{2 pt} \begin{eqnarray}
h(t)  =  0.359 & - &  0.488 \cos \left (\frac{2 \pi t}{T_0} \right) \nonumber \\  & + &  0.141 \cos \left ( \frac{4 \pi t}{T_0} \right)  - 0.012 \cos \left ( \frac{6 \pi t}{T_0} \right) 
\end{eqnarray}} for  $t \in [0, T_0]$ and $h(t) = 0$, otherwise. We choose $T_0 = 0.12$ as the final (unitless)  pulse duration obtained after combining the frequency lines.   For each measurement point, the received signal ${\bf y}$ is recorded for unitless time values 0.1--0.7, that is, 0.7--5.0 $\mu$s. The corresponding sampling rate is 15 MHz.  

The present noise estimates are based on the targeted antenna specifications of the DISCUS mission concept. We assume that the measurement errors contained by ${\bf n}$ will be mainly caused by the galactic background noise and the Sun. At $20$ MHz, the galactic noise can be estimated to be around 5E-20 W/($\hbox{m}^2$Hz). The radiation from the Sun at a distance of $1\,\text{AU}$ and 20 MHz is about 2E-19 W/($\hbox{m}^2$Hz) and 2E-23 W/($\hbox{m}^2$Hz) for its active and inactive (quiet) phase of sunspot activity, i.e., for surface temperatures 1E+6 and 1E+10 K, respectively \cite{barron1985, kraus1967}. During the active phase, the Sun emits radio-frequency waves in time scales varying from seconds to hours. As the reference level for modeling inaccuracies, we use the observation that, in CONSERT \cite{kofman2015}, the peaks related to unpredictable echoes stayed mainly $- 20$ dB below the main signal  peak. 

The total error ${\bf n}$ is assumed to be an independent Gaussian white noise term with standard deviation between -25 and 0 dB with respect to the maximal entry of the difference $|{\bf y}  - {\bf y}^{(\hbox{\scriptsize bg})}|$ between the measured and simulated signal. A Gaussian noise model is used, since ${\bf n}$ is expected to include both unknown forward  and measurement  errors, and as the sum of different independent and random error sources approaches a Gaussian distribution by the central limit theorem \cite{rice2006}. The measurement errors due to the spectral radiation flux density $F$ [W/($\hbox{m}^2$Hz)] of the galactic background and the Sun are approximated using this relative scale. We estimate the relative standard deviation of the measurement noise  with the formula
\begin{equation}
\sigma_m = \sqrt{ \frac{ F A_{\hbox{\scriptsize eff}} B_{\ell} }{ P_{TX} }},
\end{equation} 
i.e., it is the square root of the ratio between the absorbed noise and transmitted signal amplitude. With $\sigma_m$, the noise will have the desired average power given by $F$. As the antenna aperture, the standard approximation for the half-wavelength dipole antenna $A_{\hbox{\scriptsize eff}} = G  \lambda^2 / (4 \pi)$ is used. The antenna of each spacecraft  is assumed to be parallel to the orbiting plane normal and  pointed towards the Sun during the measurements (Figure \ref{bistatic_measurement}). 

\subsection{Inversion Process}

In this study, the  process of finding an estimate $\varepsilon^\ast_r$ for the relative permittivity $\varepsilon_r$, the unknown of the inverse problem, consists of the following  three stages presented in  \cite{pursiainen2016}: 
\begin{enumerate}
\item {\em Forward Simulation}. For each transmission and/or measurement position, the signal can be  propagated using the leap-frog iteration. A parallel computing cluster can be used, as the iteration is an  independent process between different source points. In this study, altogether sixteen Nvidia Tesla P100 GPUs belonging to the {\em Narvi} cluster of Tampere University of Technology were used. Running the wave simulation for a single source position took about 71  and 184 minutes in the case of  the background and exact data, respectively. The matrix--vector products including the inverse of the mass matrix \cite{pursiainen2016} were evaluated using the preconditioned conjugate gradient method with the lumped diagonal preconditioner, i.e., a matrix with row sums on its diagonal. The system matrices required a total of about 4 GB memory of which the mass matrix took 0.62 GB. The total GPU memory consumption during the forward simulation was around 10 GB, since the transposed matrices were stored separately in the memory to speed up the matrix-vector products of the iteration. 
\item {\em Linearization}. The forward model can be linearized in a coarse nested mesh \cite{pursiainen2016} covering the asteroid which allows achieving an invertible system size. To speed up the parsing of the Jacobian matrix, the Tikhonov regularized deconvolution routine applied can be formulated as a pixelwise parallel algorithm. In the linearization stage, we applied a coarse mesh of 80000 tetrahedra. Running a parallelized 32-thread version of the parsing routine took 340 and 720 seconds for the monostatic and bistatic full-angle data in a Lenovo P910 workstation equipped with two Intel Xeon E5 2697A v4 2.6 GHz 16-core processors and 128 GB of RAM. The size of the resulting Jacobian matrix  took 4.7 GB and 9.4 GB memory space, respectively. 
\item {\em Reconstruction procedure}.  
The relative permittivity distribution can be reconstructed via the total variation (TV) regularized iteration presented in \cite{pursiainen2016}.  The reconstruction was found via one iteration step using the regularization parameter values $\alpha = 0.1$ and $\beta = 0.001$ \cite{pursiainen2016}. The first one of these controls the overall regularization level and the second one the weighting ratio between the TV and norm-based regularization (the larger the value the more weight on the norm). To invert the regularized system matrix, we applied the conjugate gradient method with the stopping criterion 1E-5 for the relative residual norm. Computing a single reconstruction in the P910 workstation required around 500 and 400 conjugate gradient steps for the monostatic and bistatic full-angle system. The computation time was 100 and 170 seconds, respectively.
\end{enumerate}

\subsubsection{Inversion Accuracy}
\label{inversion_accuracy}

The accuracy of the estimate $\varepsilon^\ast_r$ obtained is measured using the relative overlap and value error (ROE and RVE), that is, the percentages {\setlength\arraycolsep{2 pt}\begin{eqnarray} \hbox{ROE} & = & 100 \left( 1 - \frac{\hbox{Volume}( \mathbf{T} \cap \mathbf{V}) }{ \hbox{Volume}(\mathbf{S})} \right) \\ \hbox{RVE} & = & 100 \left( 1 - \left|  \frac{\int_{\mathbf{T}}  (\varepsilon^\ast_r - \varepsilon_r^{(\hbox{\scriptsize bg})})  \, \hbox{d} V}{ \int_{\mathbf{T}}  (\varepsilon_r -\varepsilon_r^{(\hbox{\scriptsize bg})})  \, \hbox{d} V  } \right| \right).  \end{eqnarray}} where  $\varepsilon_r$ denotes the actual permittivity distribution and $\varepsilon_r^{(\hbox{\scriptsize bg})}$ is the background (initial guess). The set $\mathbf{S}$ is a region of interest (ROI) including both the surface layer and the voids and $\mathbf{V} = \mathbf{S} \cap \mathbf{R}$ denotes the overlap between the ROI and the set $\mathbf{R}$ in which a given reconstruction is smaller than a limit such that $\hbox{Volume}(\mathbf{R}) = \hbox{Volume}(\mathbf{S})$. The target set $\mathbf{T}$ refers to (1) the full ROI, i.e., $\mathbf{T} = \mathbf{S}$, (2) its surface part or (3) voids. 

\subsection{Discretization}

The spatial domain was discretized as presented in \cite{pursiainen2016} and Appendix, using an unstructured tetrahedral mesh with accurate interior surfaces for all the modeled structures including the asteroid and the sphere $\mathcal{S}$. The fields $\vec{E}$ and ${g}^{(i)}$, $i = 1,2,3$ were discretized using piecewise linear and (element-wise) constant finite element basis functions, respectively. The temporal interval $[0, T]$ was divided into regular subintervals. The leap-frog based finite element time-domain algorithm (FETD) \cite{li2012,schneider2016,carley2008} presented in \cite{pursiainen2016} was utilized to obtain the fields and their linearizations. The scattered far-field for the source (\ref{point_source}) was simulated as follows: 
\begin{enumerate}
\item Calculate the incident far-field on the surface $\mathcal{S}^-$
\item Solve the weak form (\ref{weak_form_7})--(\ref{weak_form_8}). 
\item Extrapolate the scattered far-field via the restriction of $E_\mathtt{S}$ to $\mathcal{S}$. 
\end{enumerate}

The surface integral terms of  (\ref{surface_source}) and (\ref{far_field_measurement_1})--(\ref{far_field_measurement_3}) were evaluated over the triangulated surface of $\mathcal{S}$ using the one point (barycenter) quadrature rule. In order to prevent numerical noise due to non-smoothness of the incident field at the boundary between $\mathcal{S}^-$  and $\mathcal{S}^+$, the following smoothed approximation for  $\langle \vec{g}_{\mathtt I}, \! v ; \mathcal{S} \rangle$ was applied:
\begin{equation}
\langle \vec{g}_{\mathtt I}, \! v ; \mathcal{S} \rangle \! \approx \! \!   \int_{\mathcal{S^-}} \! \! \! \! ( \vec{g}_{\mathtt I} \cdot \vec{n} ) v  \,  \hbox{d} S  +    \int_{\mathcal{S^+}} \! \! \! \! ( \vec{g}_{\mathtt I} \cdot \vec{n} ) v \exp \! \left( \frac{- \gamma (\vec{x} \! - \! \vec{p}) \! \cdot \! \vec{n}}{1 \! - \! (\vec{x} \! - \!  \vec{p}) \! \cdot \! \vec{n}} \right) \! \hbox{d} S
\end{equation}
with the parameter $\gamma=10$ determining the decay rate for the second term. The shortest distance between  $\vec{p}$ and $\mathcal{S}$  was utilized as a time shift  to cancel out the signal travel-time between $\vec{p}$ and $\mathcal{S}$ in the simulated  data sequence in a systematic way. 

\subsubsection{Domain}

\begin{figure}
\begin{center}
\begin{minipage}{3.9cm}
\includegraphics[width=3.5cm]{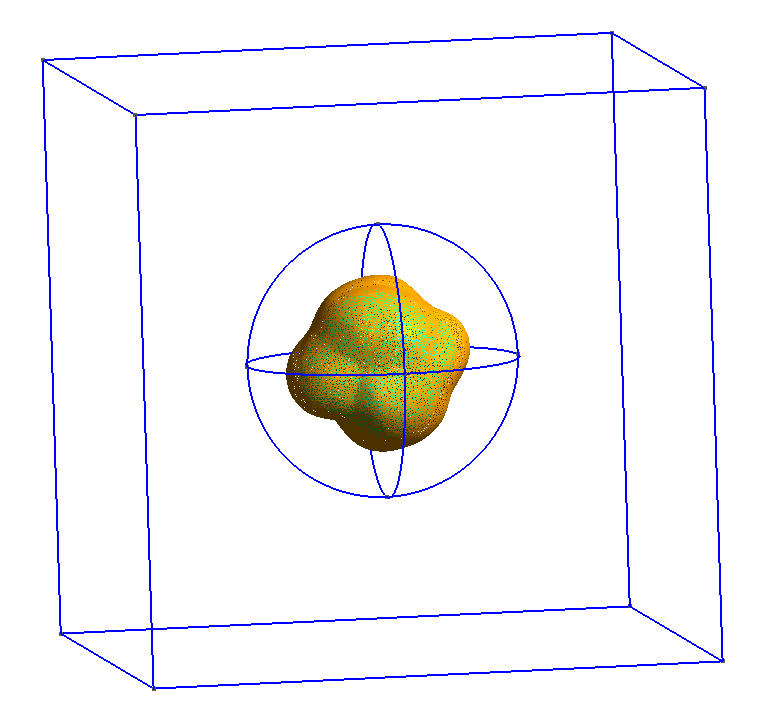}
\end{minipage}
\begin{minipage}{3.9cm}
\includegraphics[width=3.5cm]{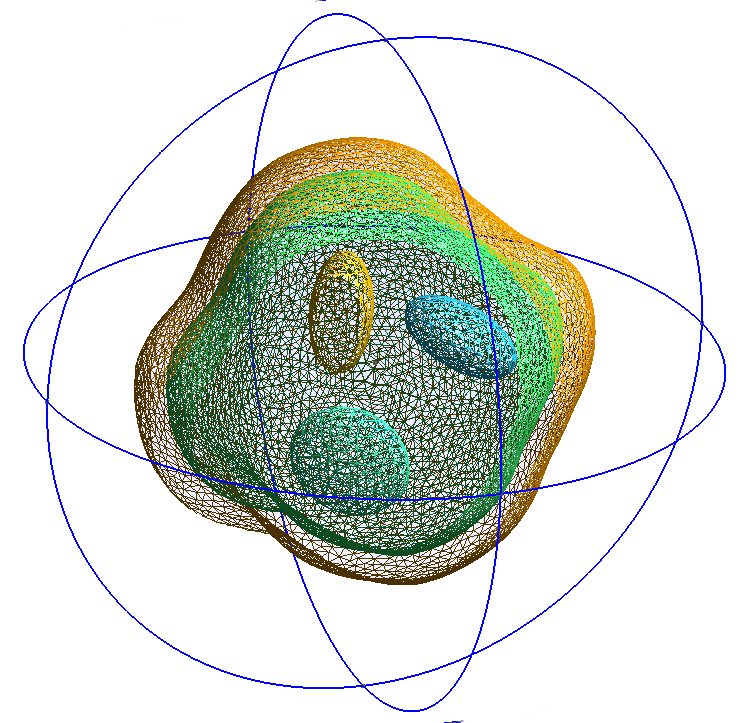}
\end{minipage}
\end{center}
\caption{The present test domain $\Omega = \Omega_1 \cup \Omega_2$ (left) is an origin centric cube. The interior of the sphere $\mathcal{S}$ centered at origin formed the subdomain $\Omega_2$ (right) the exterior of which $\Omega_1$  contained a split-field perfectly matched layer \cite{pursiainen2016, schneider2016} to simulate open field scattering. As the test target, we utilized the surface model (12260 triangles) of asteroid 1998 KY26 which can be associated through scaling  with  a 550 m diameter asteroid  (Table \ref{scaling}). The volumetric finite element discretization used in the reconstruction procedure was created based on the multi-layer surface mesh model illustrated in the pictures. \label{domain_fig}}
\end{figure}

The present test domain $\Omega = \Omega_1 \cup \Omega_2$, is an origin centric  cube (Figure \ref{domain_fig}). The interior of the sphere $\mathcal{S}$ centered at origin formed the subdomain $\Omega_2$ the exterior of which $\Omega_1$ contained a split-field perfectly matched layer \cite{pursiainen2016, schneider2016} to simulate open field scattering. As the target, we utilized the surface model of asteroid 1998 KY26 which can be associated through scaling with a rubble pile asteroid  (Table \ref{scaling}). 

Two different conforming tetrahedral finite element meshes were used for simulating background and exact data, i.e., ${\bf y}^{(\hbox{\scriptsize bg})}$ and ${\bf y}$, respectively. Each one consisted of a total of 14.9 M elements of which 5.3 M were contained by the asteroid. All the surfaces present in the model were modeled accurately as triangular finite element mesh boundaries. The interior structure was homogeneous in the case of the background. The exact model included the following inhomogeneities to be reconstructed: (i) a surface layer together and (ii) three deep interior anomalies. Two different meshes were applied  in order to avoid overly good data fit, i.e., the {\em inverse crime} \cite{colton1998}. The wave was propagated from zero time to the (unitless) value $T = 0.7$ (5.0 $\mu$s) using the FETD method. As the time increment  for the  background and exact asteroid mesh we employed the (unitless) values $\Delta t =$ 1E-4 and $\Delta t =$ 3.7E-5, respectively.

\begin{figure}[t] 
\begin{scriptsize} 
\begin{center} 
\begin{minipage}{8.0cm}  
\begin{center} {\bf Overall} ROE and RVE for (A). \\ \vskip0.2cm
\begin{minipage}{3.7cm}
\rotatebox[origin=c]{90}{ROE in percents (\%)} 
\begin{minipage}{3.3cm} \begin{center}  
\includegraphics[height=3.1cm]{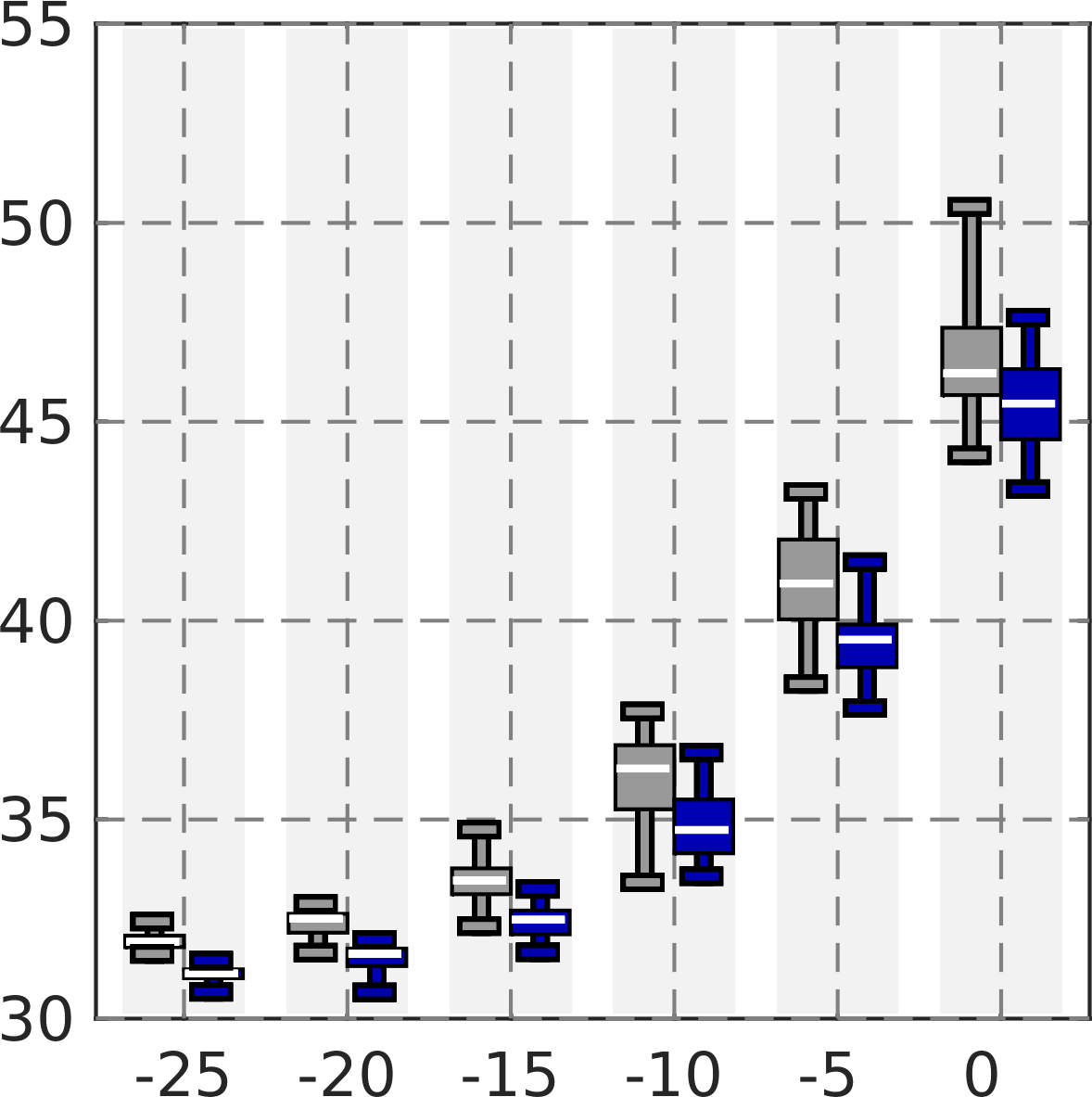} \\ Total Noise (dB)   
\end{center} 
\end{minipage}
\end{minipage}
\begin{minipage}{3.7cm}
\rotatebox[origin=c]{90}{RVE in percents (\%)} 
\begin{minipage}{3.3cm}
\begin{center}
\includegraphics[height=3.1cm]{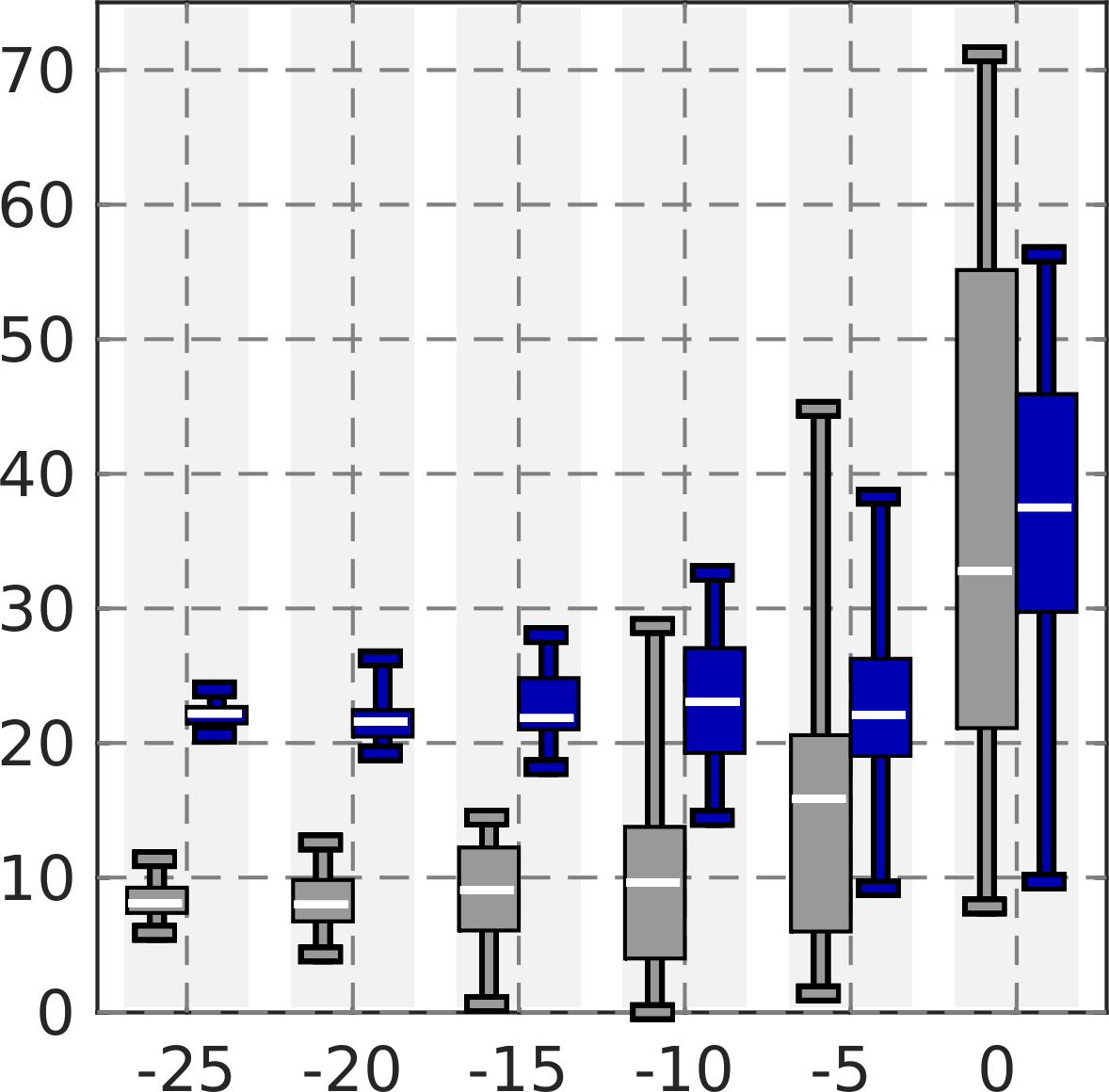}  \\ Total Noise (dB) 
 \end{center}
\end{minipage}
\end{minipage}
\end{center}
\end{minipage} 
\\ \vskip0.3cm
\begin{minipage}{8.0cm}
\begin{center}  {\bf Overall} ROE and RVE for (B). \\ \vskip0.2cm
\begin{minipage}{3.7cm}
\rotatebox[origin=c]{90}{ROE in percents (\%)} 
\begin{minipage}{3.3cm} \begin{center}  
\includegraphics[height=3.1cm]{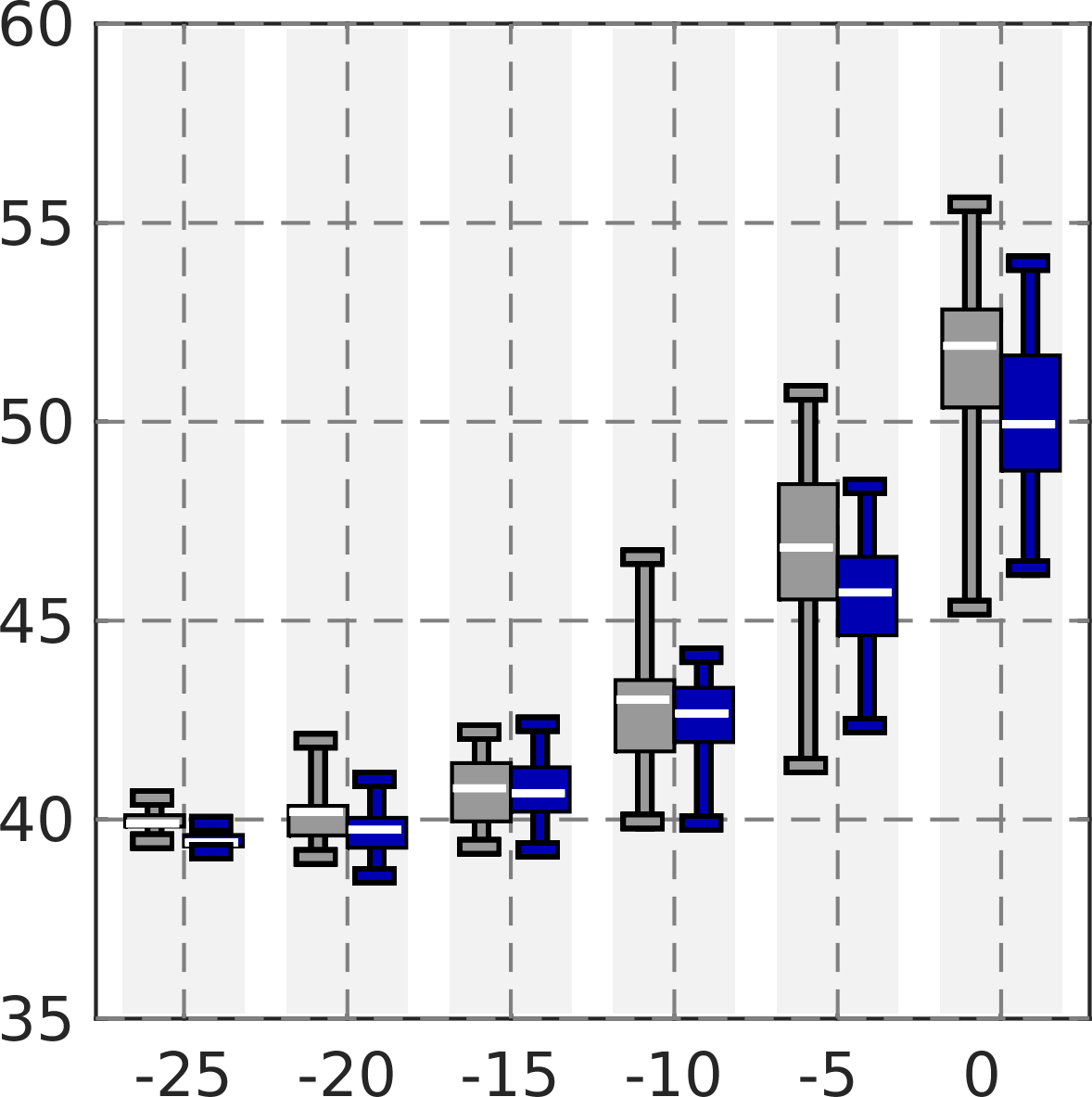} \\ Total Noise (dB)   \end{center} 
\end{minipage}
\end{minipage}
\begin{minipage}{3.7cm}
\rotatebox[origin=c]{90}{RVE in percents (\%)} 
\begin{minipage}{3.3cm}\begin{center}
\includegraphics[height=3.1cm]{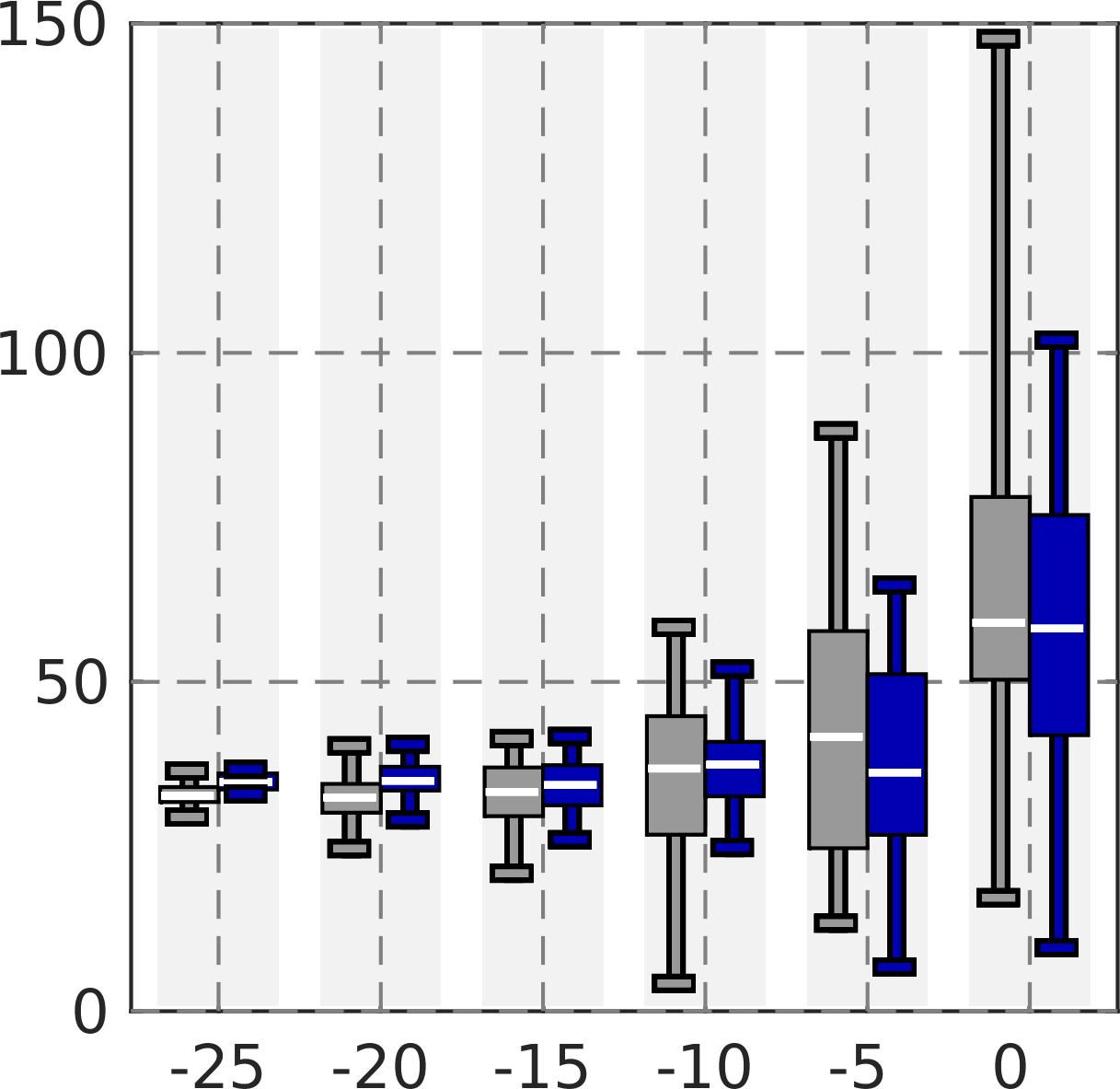}  \\ Total Noise (dB) 
 \end{center}
\end{minipage}
\end{minipage} 
\end{center}
\end{minipage} 
\\ \vskip0.3cm
\begin{minipage}{8.0cm} 
\begin{center}  {\bf Overall} ROE and RVE for (C).  \\ \vskip0.2cm
\begin{minipage}{3.7cm}
\rotatebox[origin=c]{90}{ROE in percents (\%)} 
\begin{minipage}{3.3cm} \begin{center}  
\includegraphics[height=3.1cm]{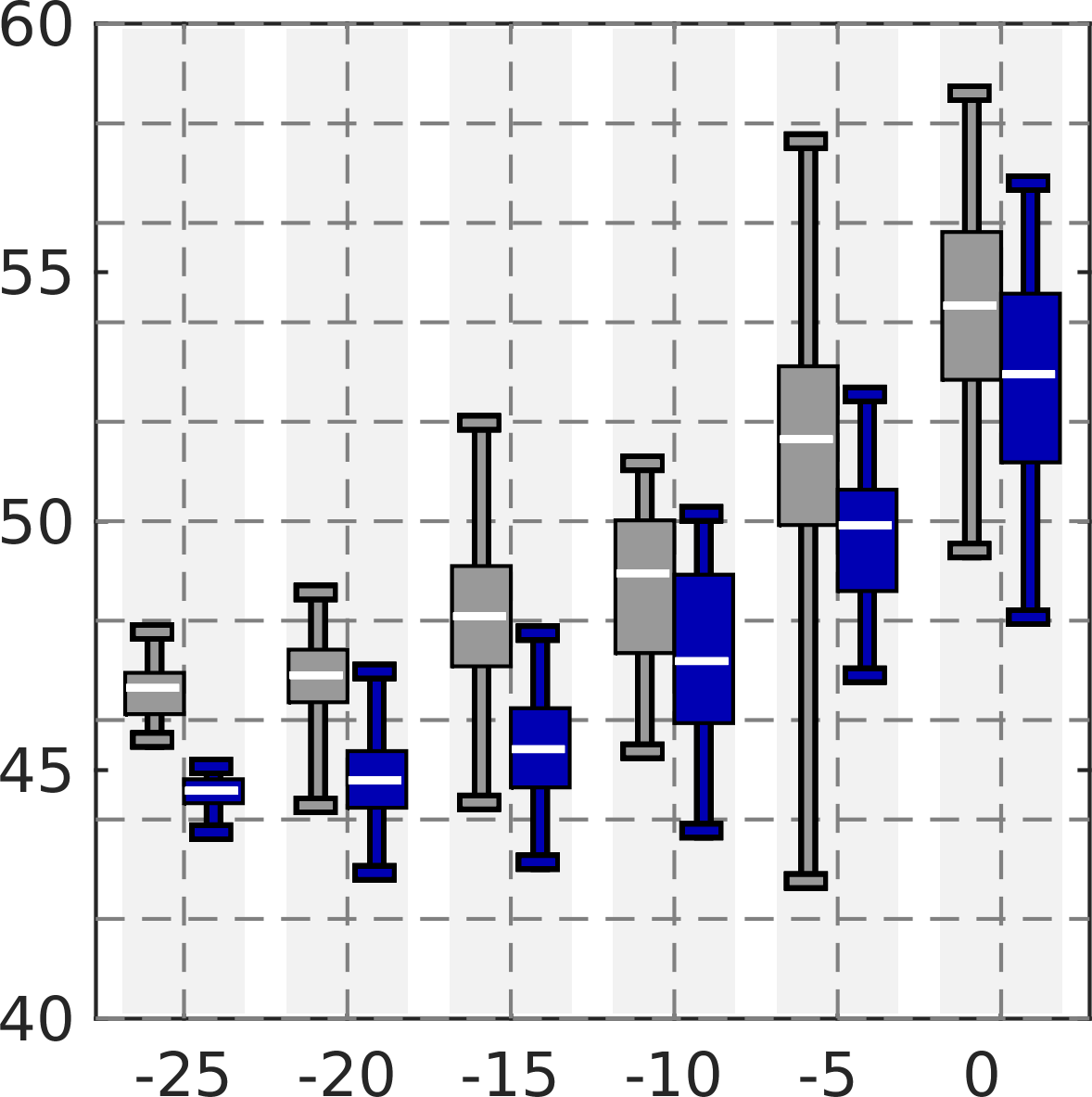} \\ Total Noise (dB)   \end{center} 
\end{minipage}
\end{minipage}
\begin{minipage}{3.7cm}
\rotatebox[origin=c]{90}{RVE in percents (\%)} 
\begin{minipage}{3.3cm}\begin{center}
\includegraphics[height=3.1cm]{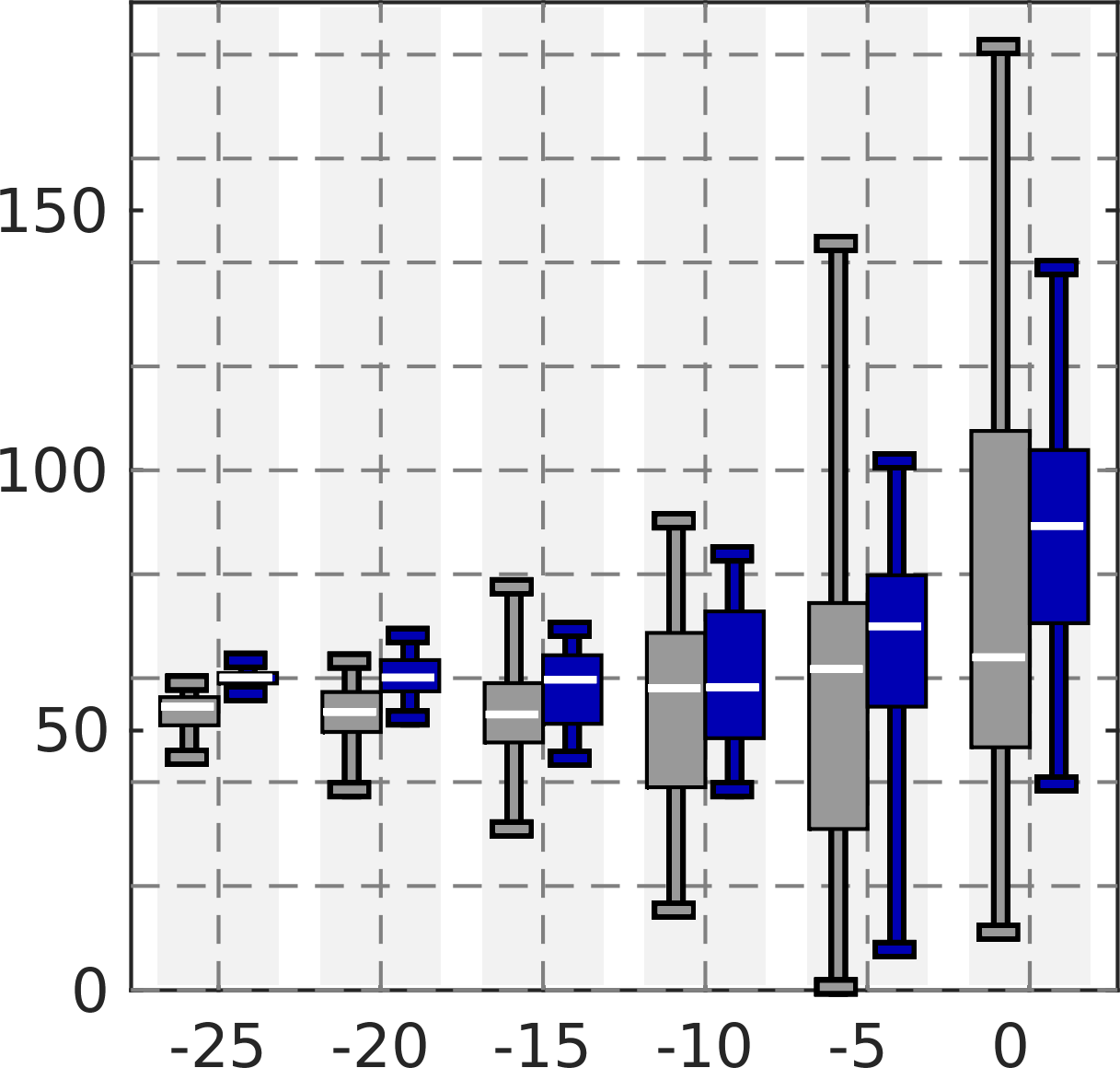}  \\ Total Noise (dB) 
 \end{center}
\end{minipage}
\end{minipage} 
\end{center}
\end{minipage} 
\end{center}
\end{scriptsize}
\caption{The {\bf overall} relative overlap and value error (ROE and RVE) in percents (\%)  in the full ROI $\mathbf{S}$. The results for the monostatic and bistatic data correspond to the light grey and dark blue box plot bars, respectively. Each bar is based on a sample of 30  reconstructions obtained with independent realizations of the total noise vector. The narrow part visualizes the interval between the minimum and maximum value in the sample. The thick part is called the interquartile range (IQR)  or spread, that is, the interval between the 25 \% and 75 \% quantile. The white line shows the median.  
\label{results_total}}
\end{figure}

\begin{figure}
\begin{scriptsize}
\begin{center}
\begin{minipage}{8.0cm}  
\begin{center} {\bf Surface} ROE and RVE for (A). \\ \vskip0.2cm
\begin{minipage}{3.7cm}
\rotatebox[origin=c]{90}{ROE in percents (\%)} 
\begin{minipage}{3.3cm} \begin{center}  
\includegraphics[height=3.1cm]{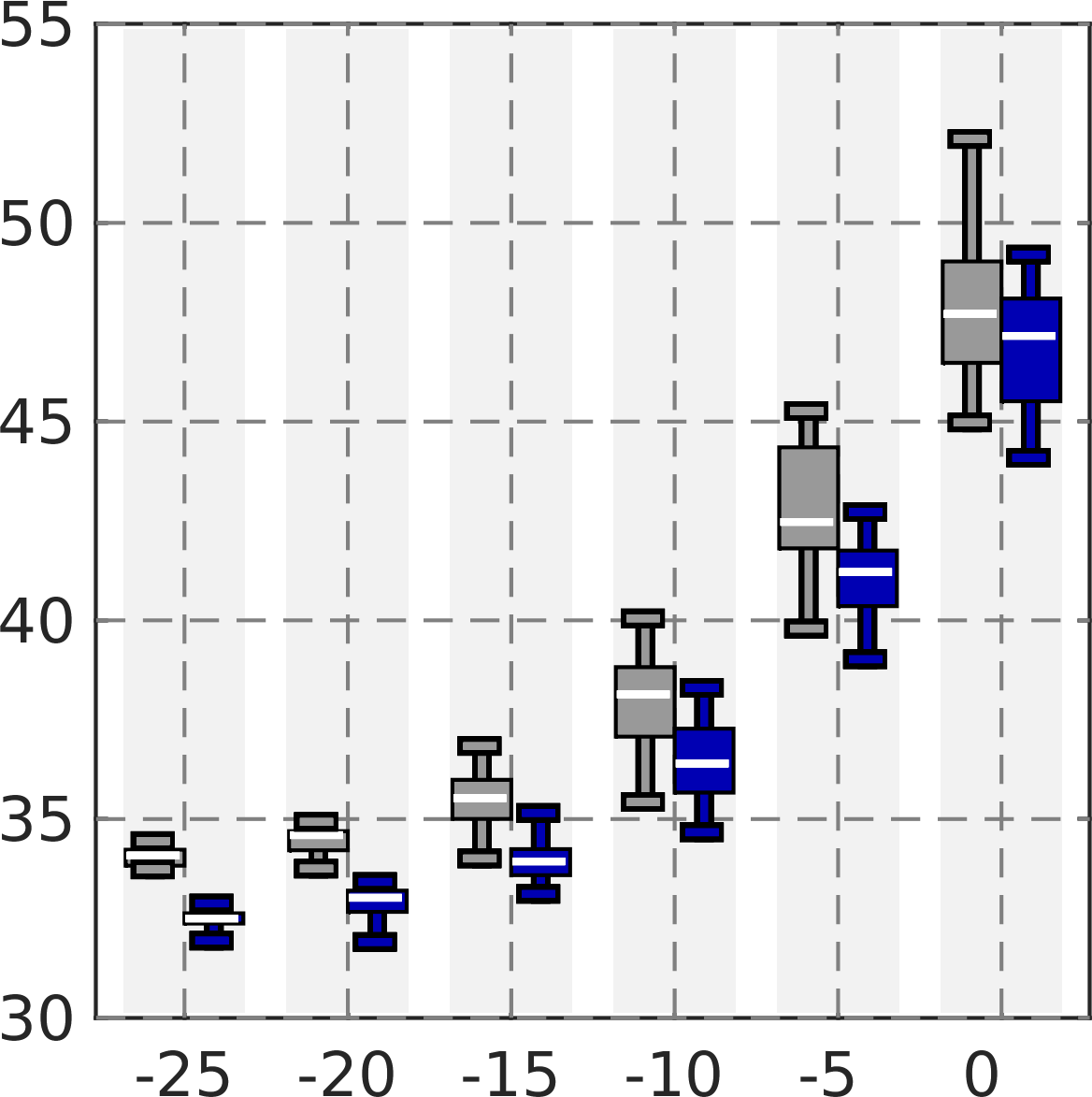} \\ Total Noise (dB)   
\end{center} 
\end{minipage}
\end{minipage}
\begin{minipage}{3.7cm}
\rotatebox[origin=c]{90}{RVE in percents (\%)} 
\begin{minipage}{3.3cm}
\begin{center}
\includegraphics[height=3.1cm]{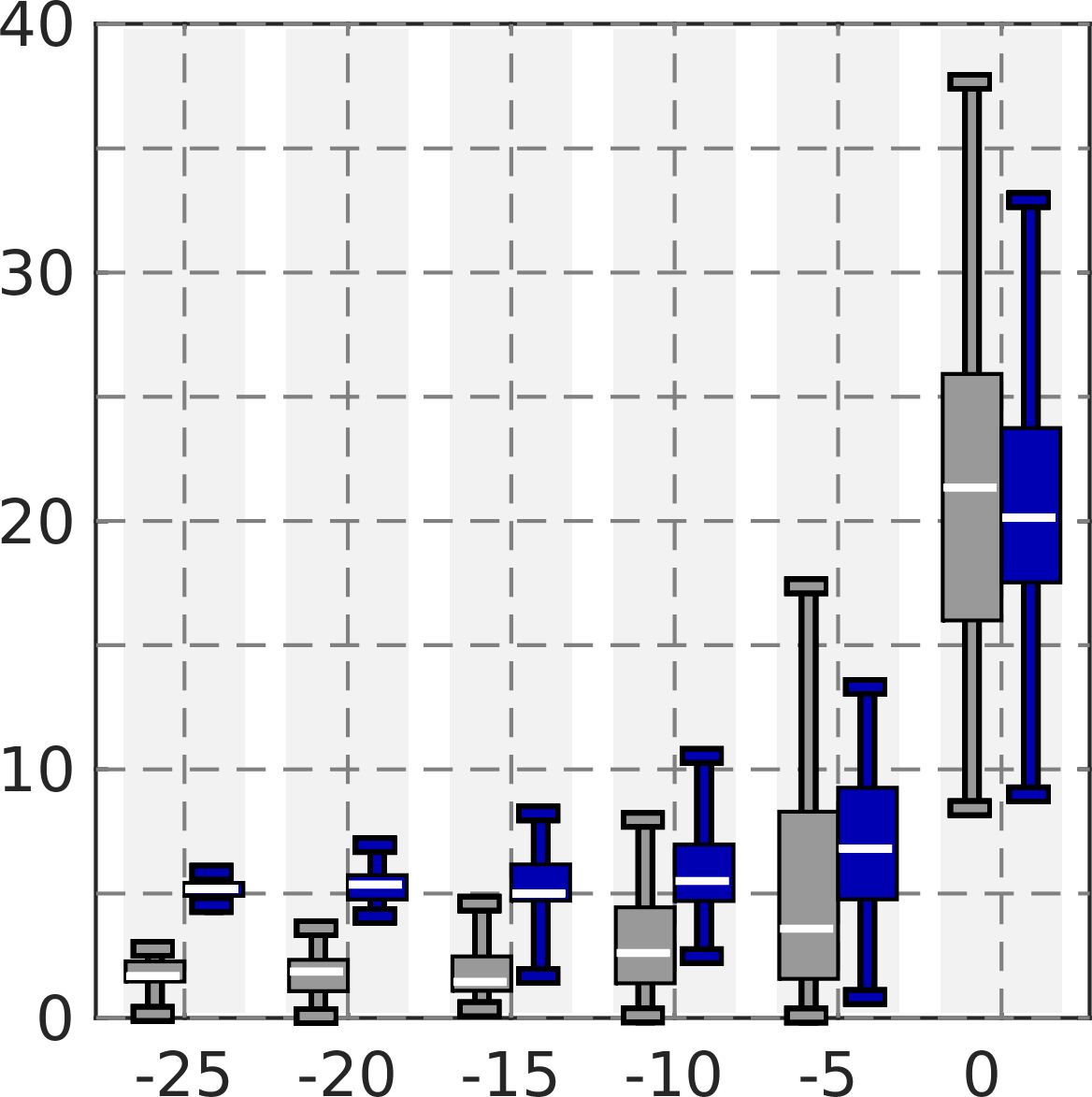}  \\ Total Noise (dB) 
 \end{center}
\end{minipage}
\end{minipage}
\end{center}
\end{minipage} 
\\ \vskip0.3cm
\begin{minipage}{8.0cm}
\begin{center}  {\bf Surface} ROE and RVE for (B). \\ \vskip0.2cm
\begin{minipage}{3.7cm}
\rotatebox[origin=c]{90}{ROE in percents (\%)} 
\begin{minipage}{3.3cm} \begin{center}  
\includegraphics[height=3.1cm]{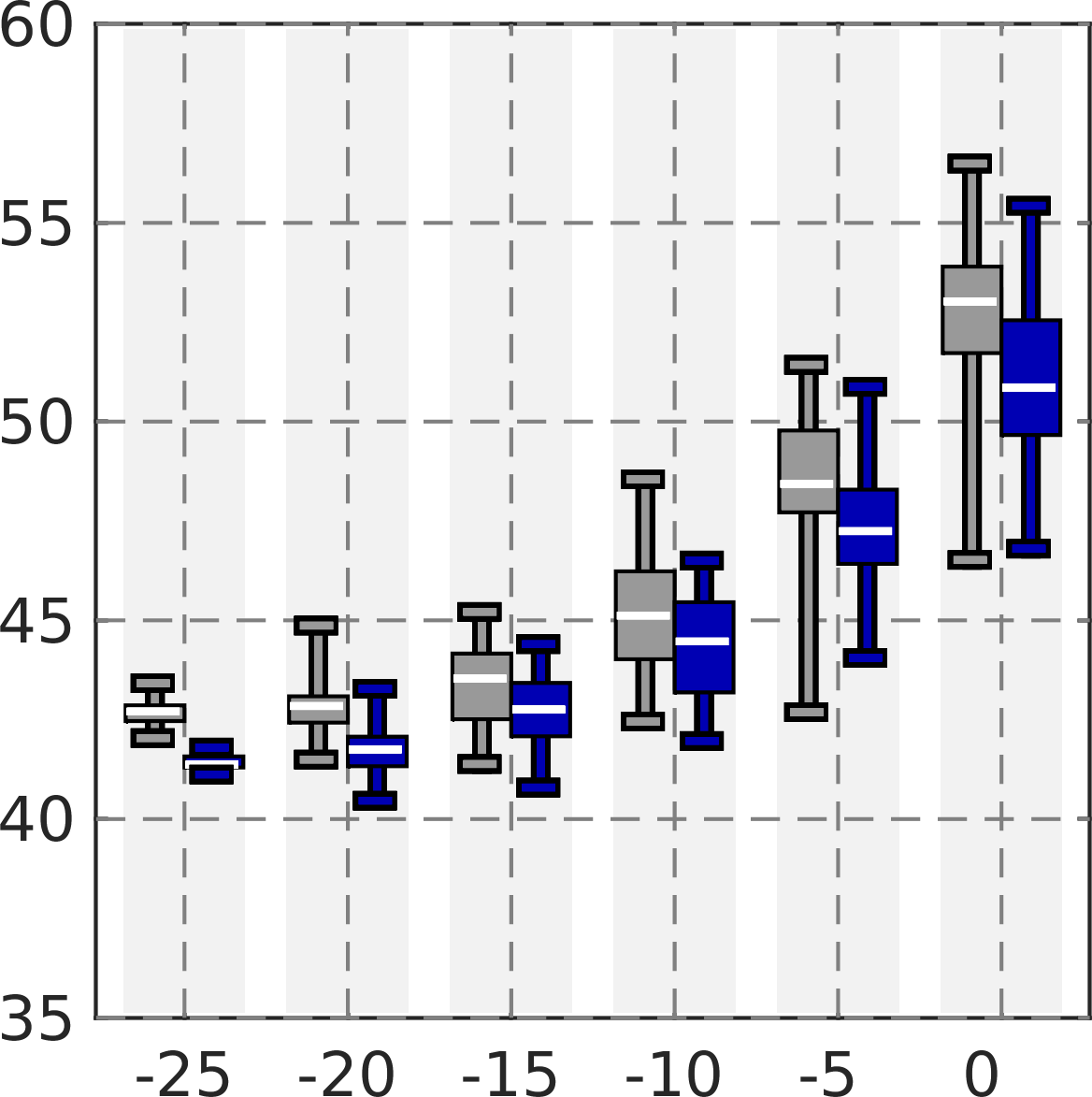} \\ Total Noise (dB)   \end{center} 
\end{minipage}
\end{minipage}
\begin{minipage}{3.7cm}
\rotatebox[origin=c]{90}{RVE in percents (\%)} 
\begin{minipage}{3.3cm}\begin{center}
\includegraphics[height=3.1cm]{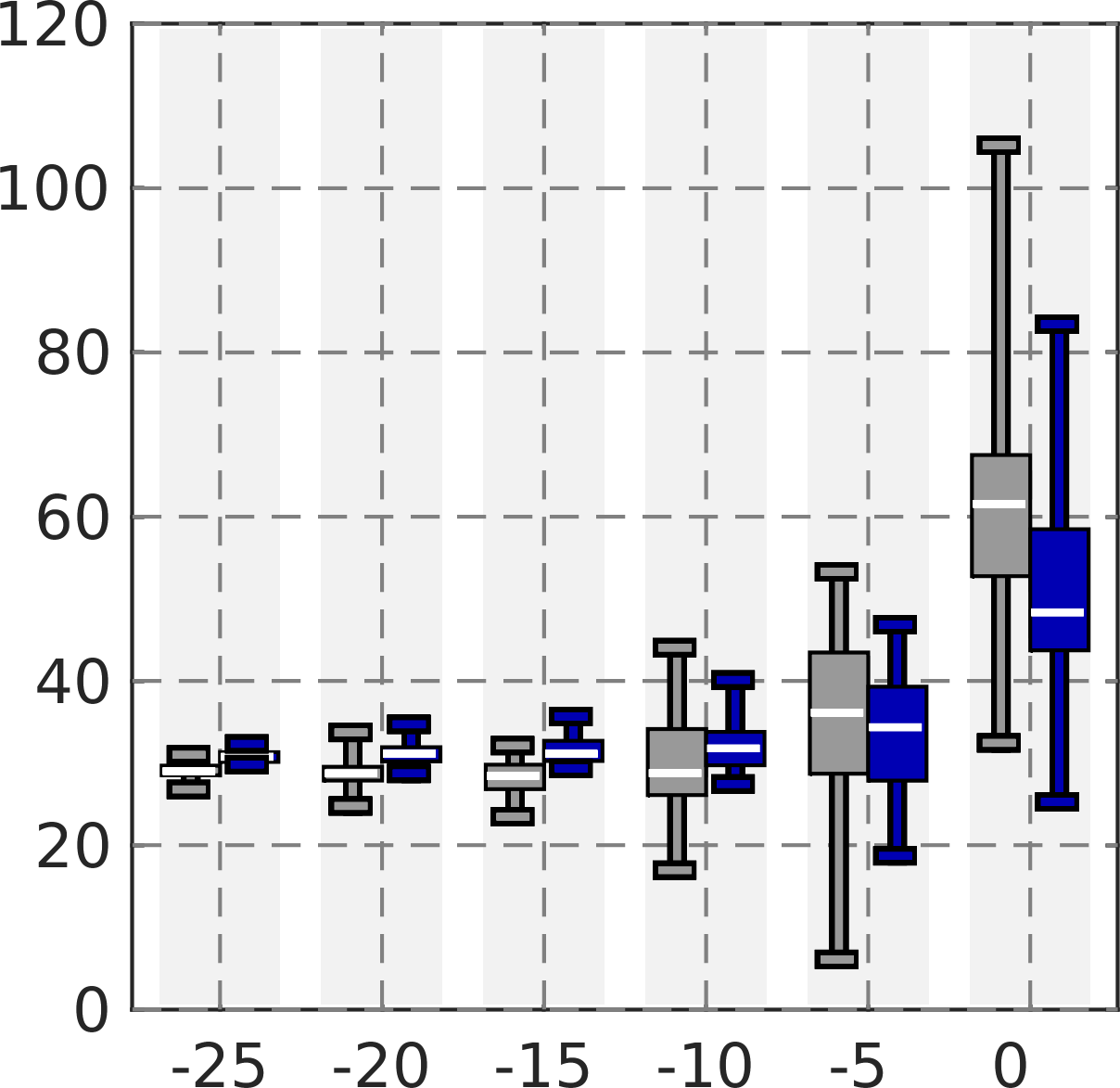}  \\ Total Noise (dB) 
 \end{center}
\end{minipage}
\end{minipage} 
\end{center}
\end{minipage} 
\\ \vskip0.3cm
\begin{minipage}{8.0cm} 
\begin{center}  {\bf Surface} ROE and RVE for (C).  \\ \vskip0.2cm
\begin{minipage}{3.7cm}
\rotatebox[origin=c]{90}{ROE in percents (\%)} 
\begin{minipage}{3.3cm} \begin{center}  
\includegraphics[height=3.1cm]{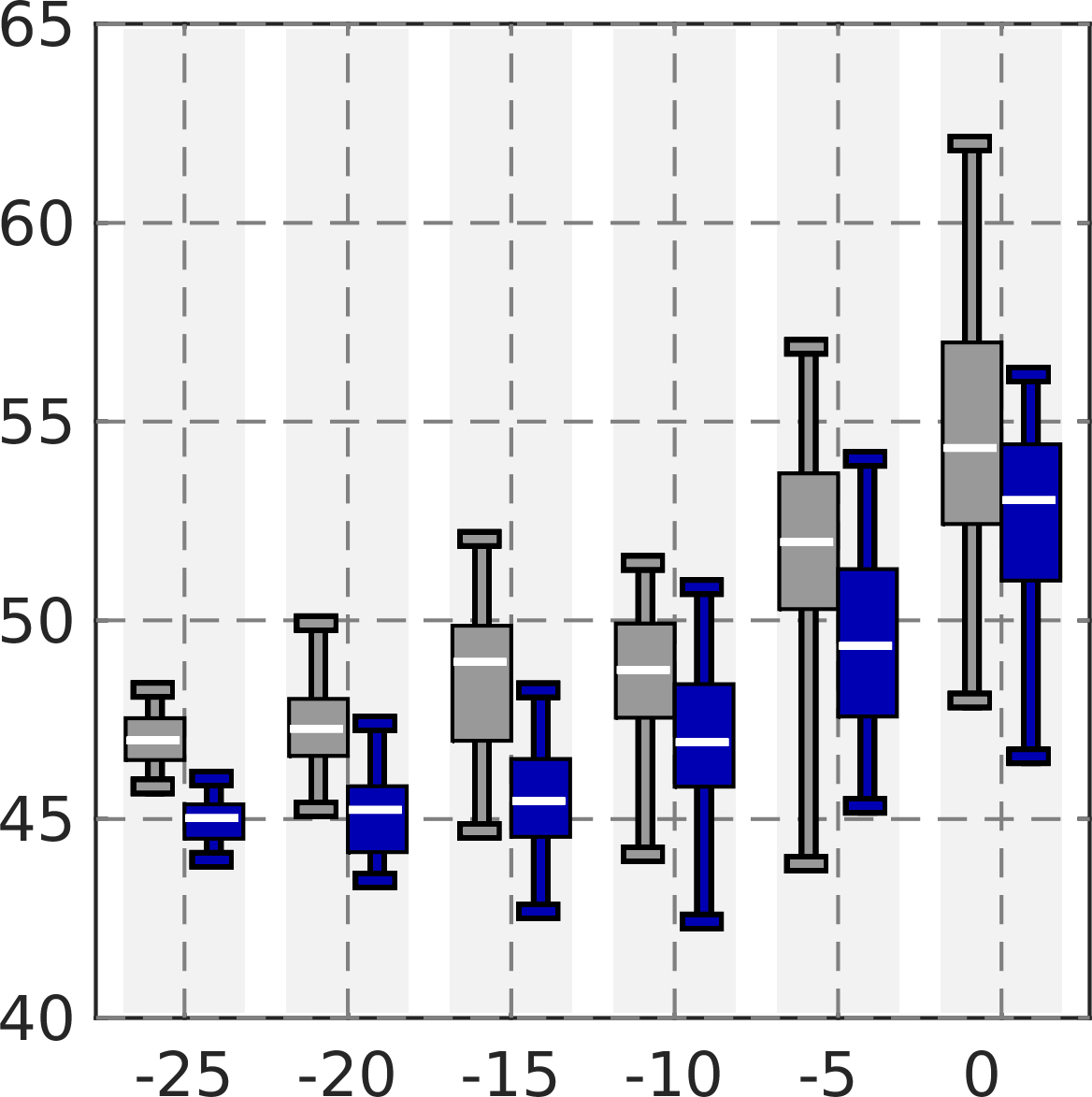} \\ Total Noise (dB)   \end{center} 
\end{minipage}
\end{minipage}
\begin{minipage}{3.7cm}
\rotatebox[origin=c]{90}{RVE in percents (\%)} 
\begin{minipage}{3.3cm}\begin{center}
\includegraphics[height=3.1cm]{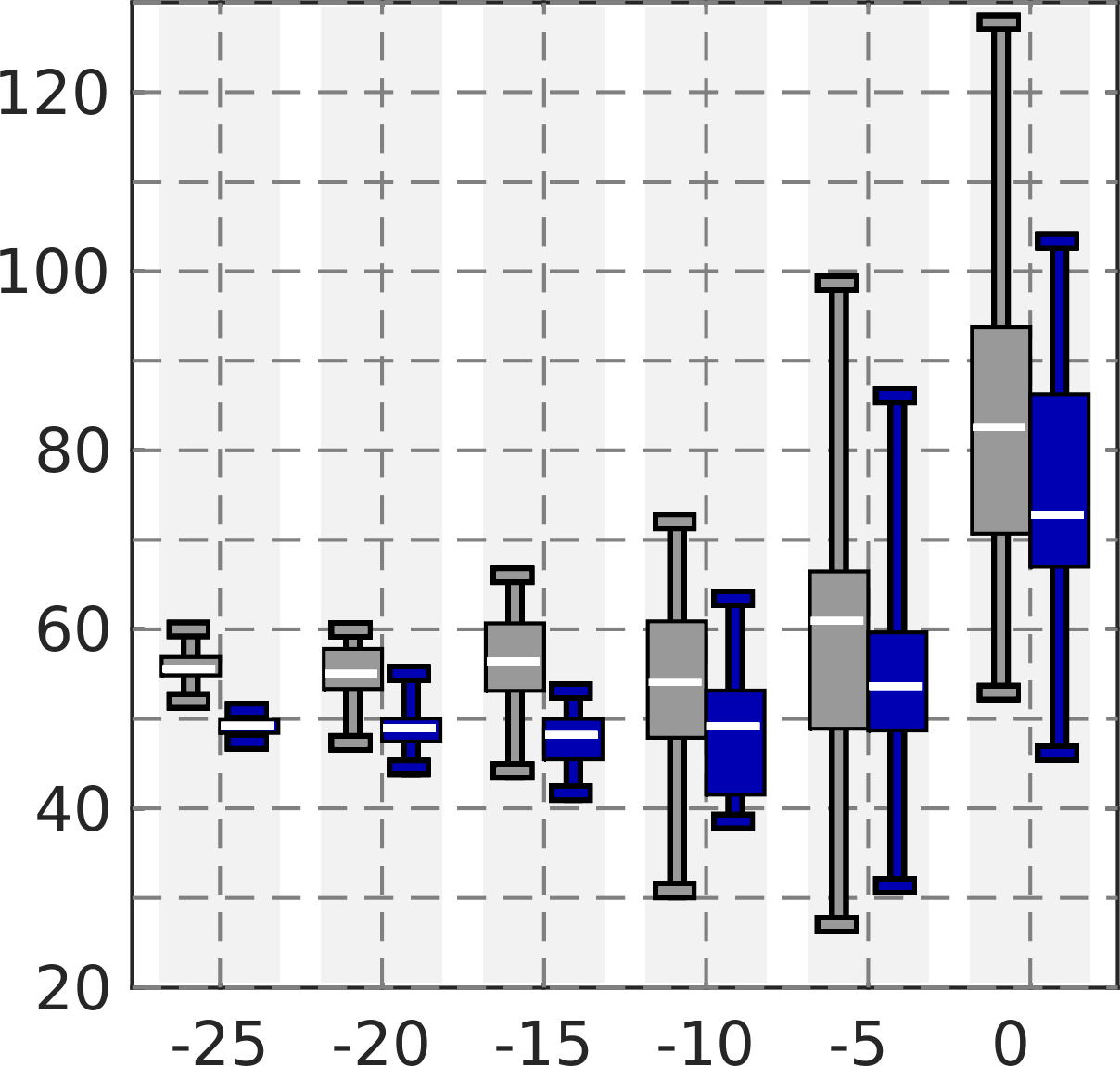}  \\ Total Noise (dB) 
 \end{center}
\end{minipage}
\end{minipage}
\end{center}
\end{minipage}
\end{center}
\end{scriptsize}
\caption{The  relative overlap and value error (ROE and RVE) in percents (\%)  for the {\bf surface part} of  ROI $\mathbf{S}$. The results for the monostatic and bistatic data correspond to the light grey and dark blue box plot bars, respectively.  
\label{results_surf}}
\end{figure}

\begin{figure}
\begin{scriptsize}
\begin{center}
\begin{minipage}{8.0cm}  
\begin{center} {\bf Deep} ROE and RVE for (A). \\ \vskip0.2cm
\begin{minipage}{3.7cm}
\rotatebox[origin=c]{90}{ROE in percents (\%)} 
\begin{minipage}{3.3cm} \begin{center}  
\includegraphics[height=3.1cm]{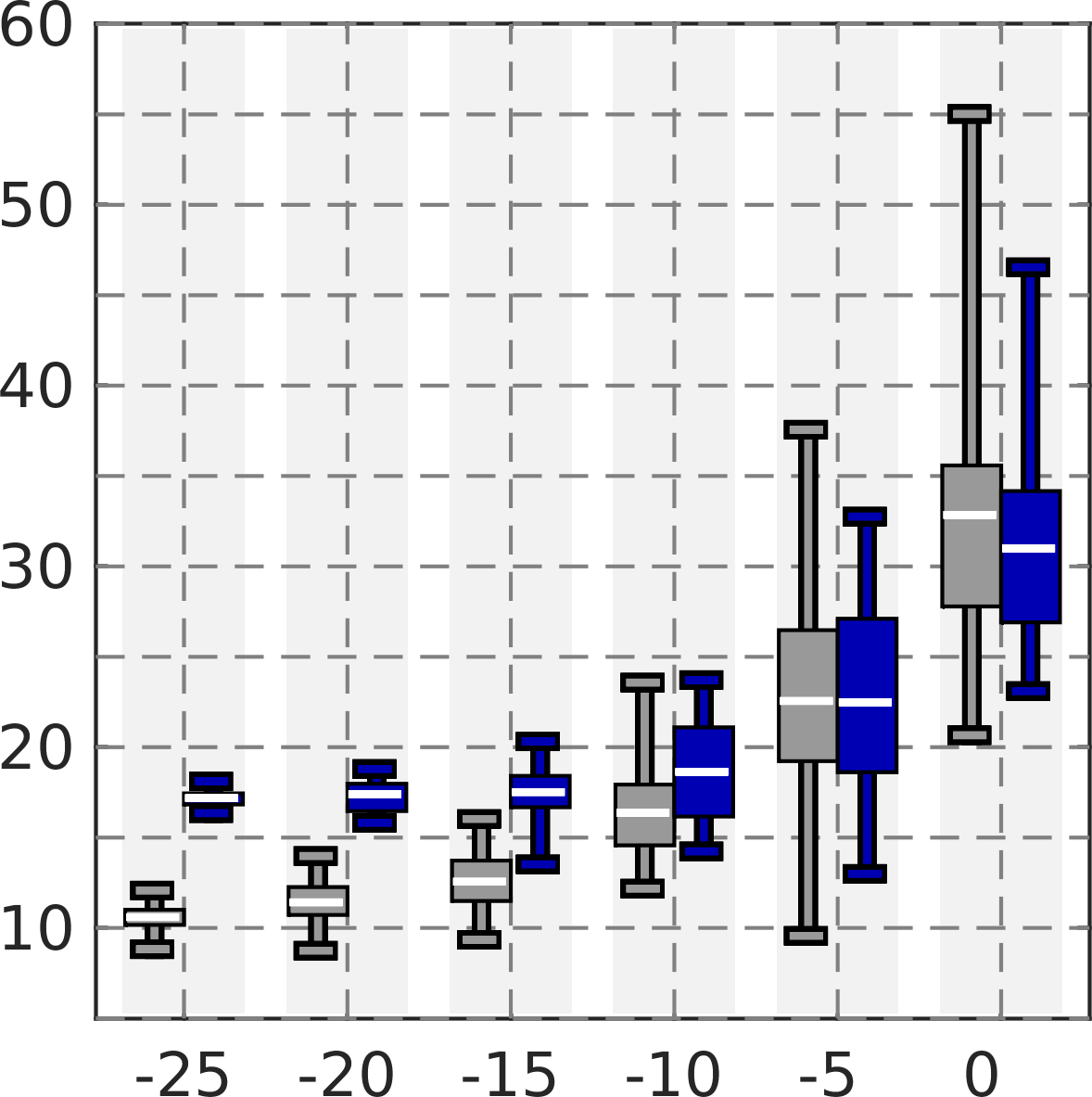} \\ Total Noise (dB)   
\end{center} 
\end{minipage}
\end{minipage}
\begin{minipage}{3.7cm}
\rotatebox[origin=c]{90}{RVE in percents (\%)} 
\begin{minipage}{3.3cm}
\begin{center}
\includegraphics[height=3.1cm]{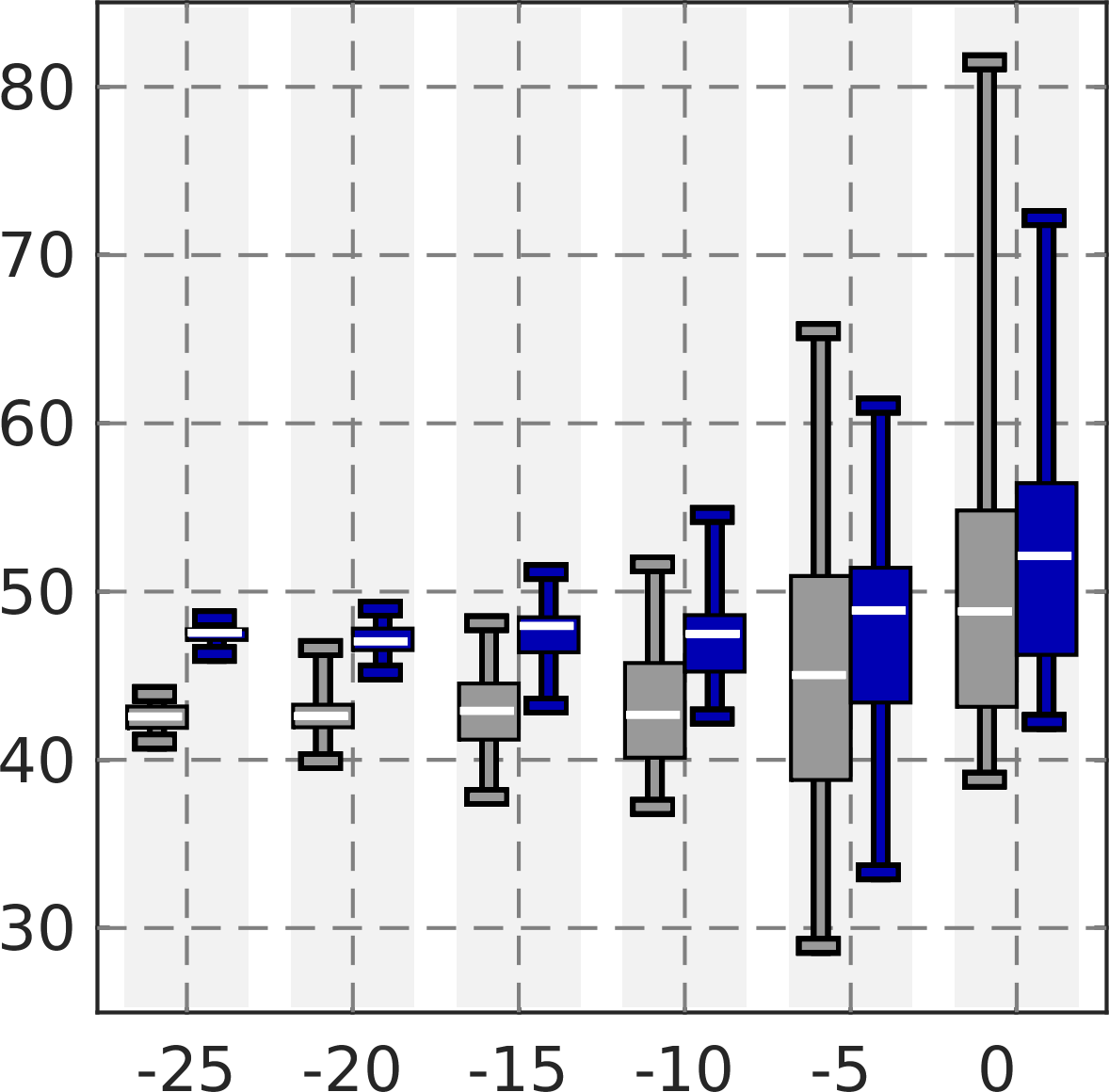}  \\ Total Noise (dB) 
 \end{center}
\end{minipage}
\end{minipage}
\end{center}
\end{minipage} 
\\ \vskip0.3cm
\begin{minipage}{8.0cm}
\begin{center}  {\bf Deep} ROE and RVE for (B). \\ \vskip0.2cm
\begin{minipage}{3.7cm}
\rotatebox[origin=c]{90}{ROE in percents (\%)} 
\begin{minipage}{3.3cm} \begin{center}  
\includegraphics[height=3.1cm]{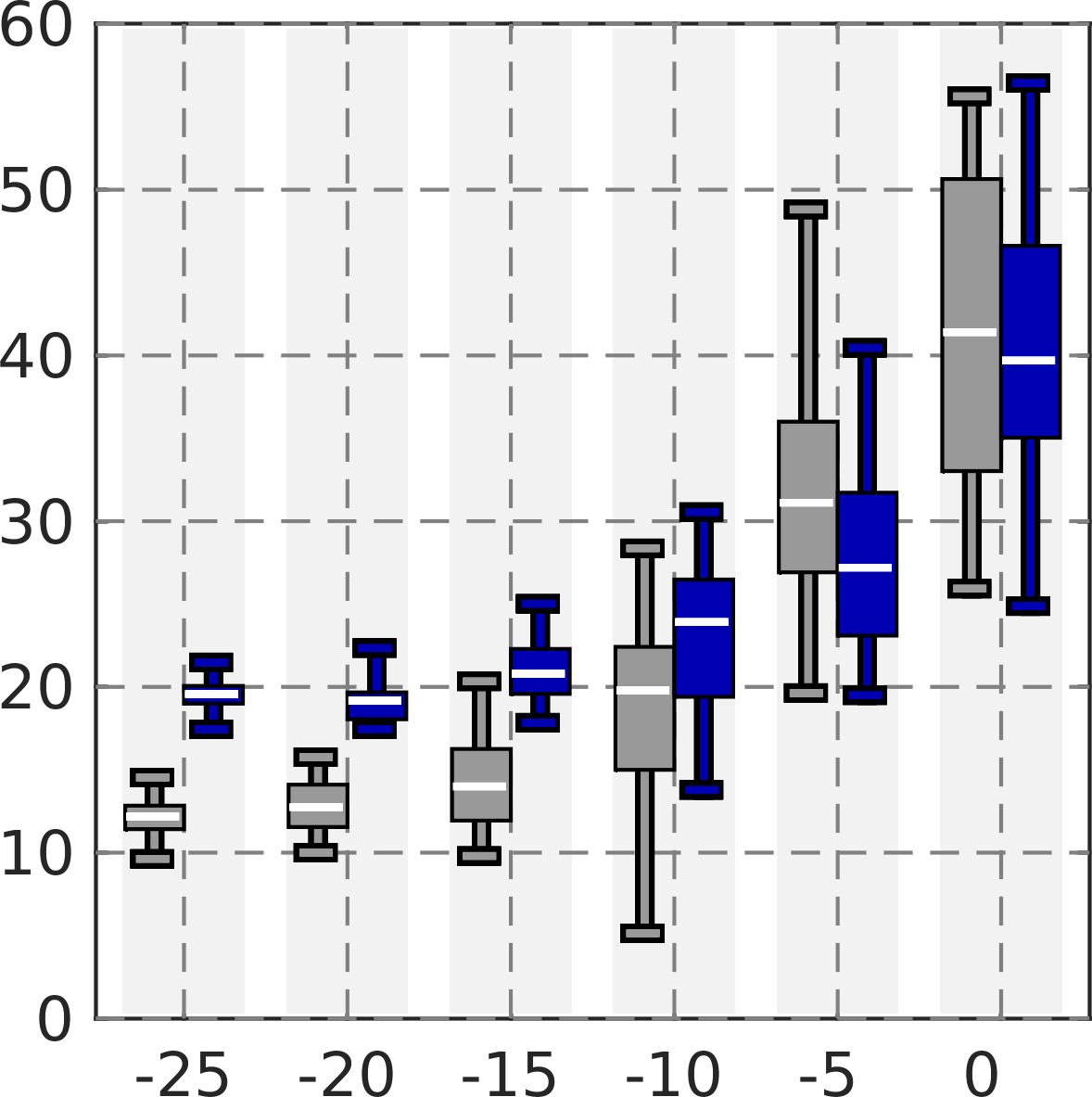} \\ Total Noise (dB)   \end{center} 
\end{minipage}
\end{minipage}
\begin{minipage}{3.7cm}
\rotatebox[origin=c]{90}{RVE in percents (\%)} 
\begin{minipage}{3.3cm}\begin{center}
\includegraphics[height=3.1cm]{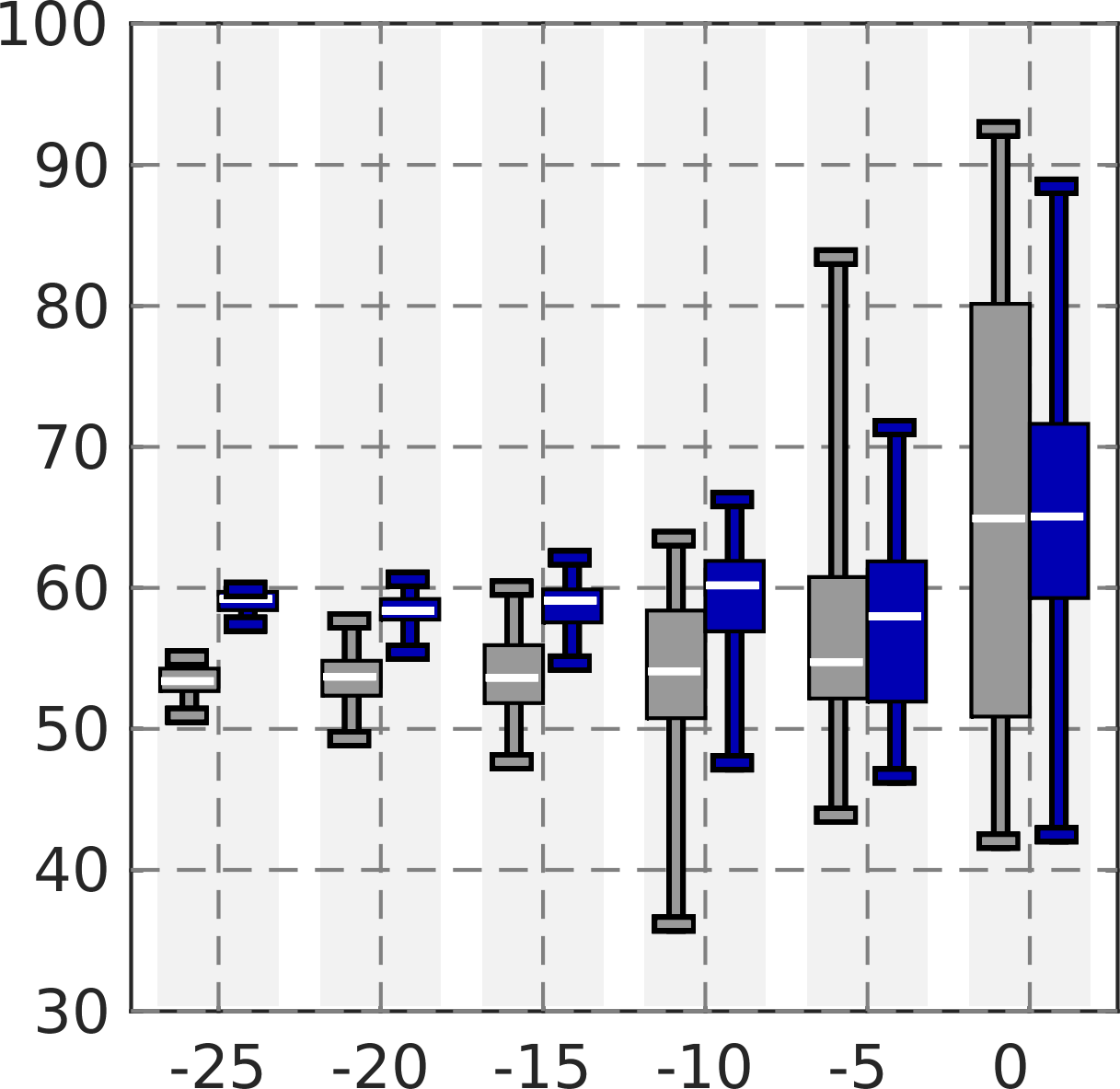}  \\ Total Noise (dB) 
 \end{center}
\end{minipage}
\end{minipage} 
\end{center}
\end{minipage} 
\\ \vskip0.3cm
\begin{minipage}{8.0cm} 
\begin{center}  {\bf Deep} ROE and RVE for (C).  \\ \vskip0.2cm
\begin{minipage}{3.7cm}
\rotatebox[origin=c]{90}{ROE in percents (\%)} 
\begin{minipage}{3.3cm} \begin{center}  
\includegraphics[height=3.1cm]{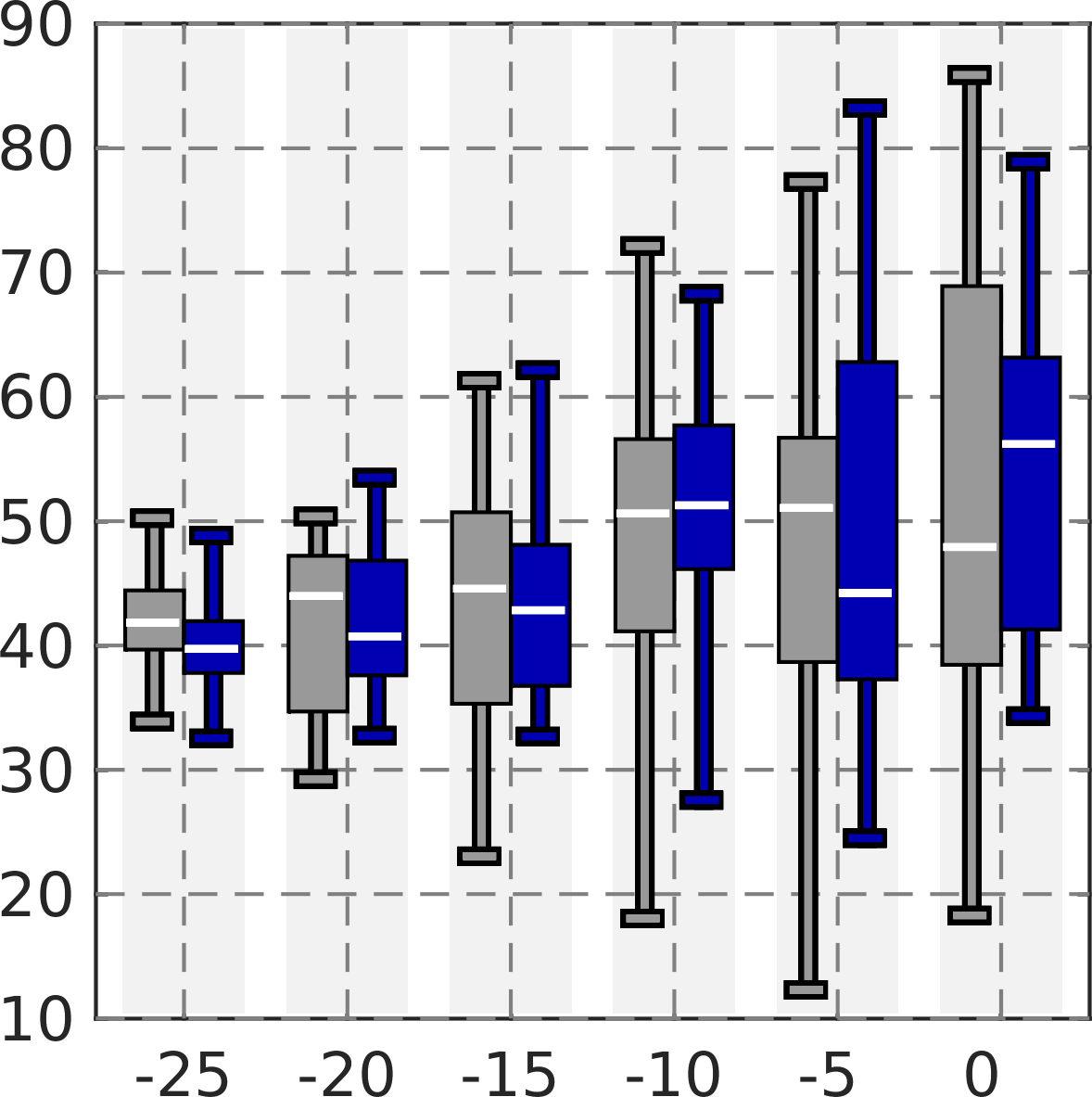} \\ Total Noise (dB)   \end{center} 
\end{minipage}
\end{minipage}
\begin{minipage}{3.7cm}
\rotatebox[origin=c]{90}{RVE in percents (\%)} 
\begin{minipage}{3.3cm}\begin{center}
\includegraphics[height=3.1cm]{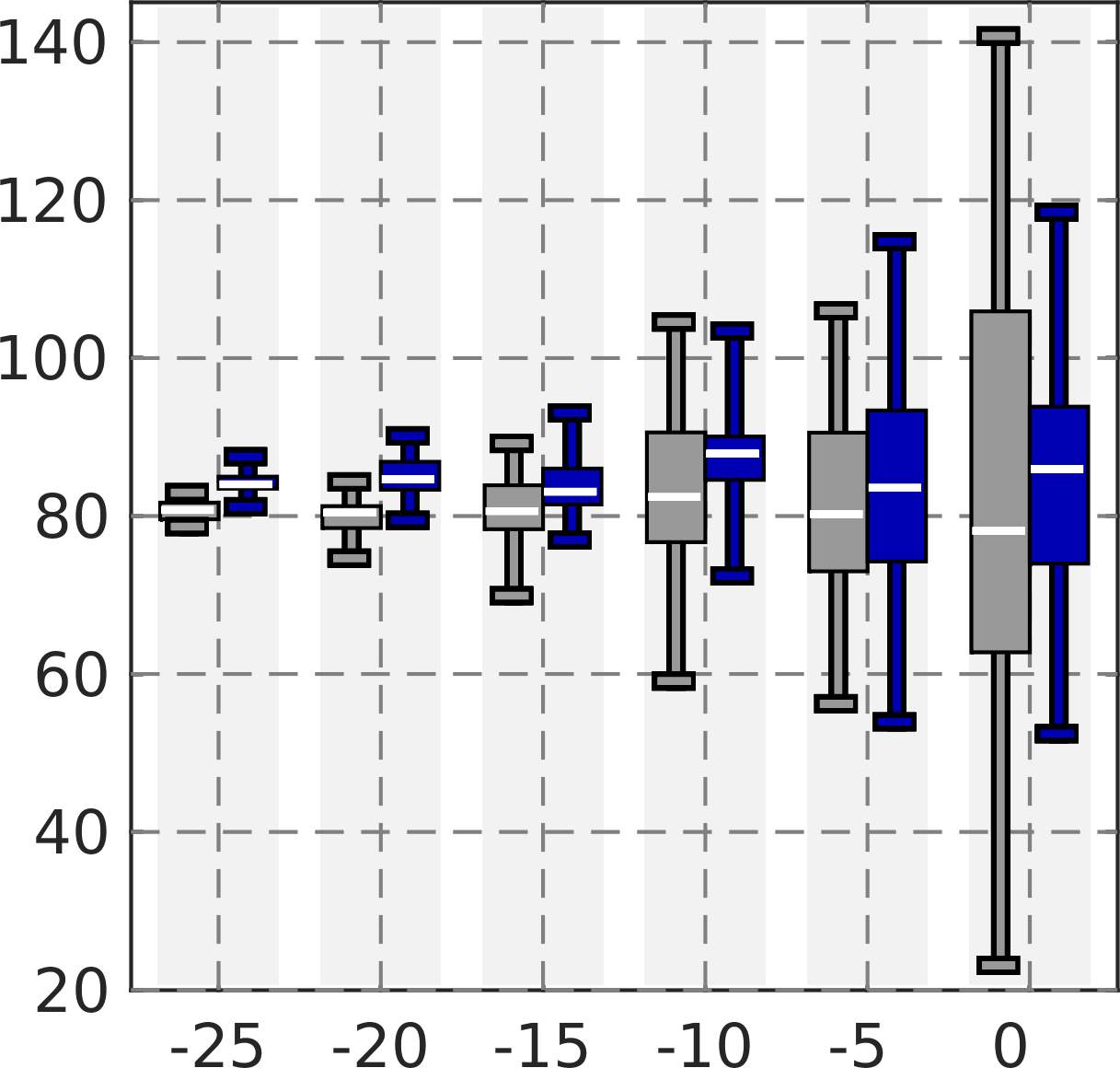}  \\ Total Noise (dB) 
 \end{center}
\end{minipage}
\end{minipage} 
\end{center}
\end{minipage}
\end{center}
\end{scriptsize}
\caption{The  relative overlap and value error (ROE and RVE) in percents (\%)  for the {\bf deep interior} part  of  ROI $\mathbf{S}$. The results for the monostatic and bistatic data correspond to the light grey and dark blue box plot bars, respectively.  
\label{results_deep}}
\end{figure}

\begin{figure*} \begin{framed} \begin{framed}
\begin{center} \begin{scriptsize} 
Clipping plane: $z = 0$ \\
\begin{minipage}{3.5cm} \begin{center}  
\includegraphics[width=3.0cm]{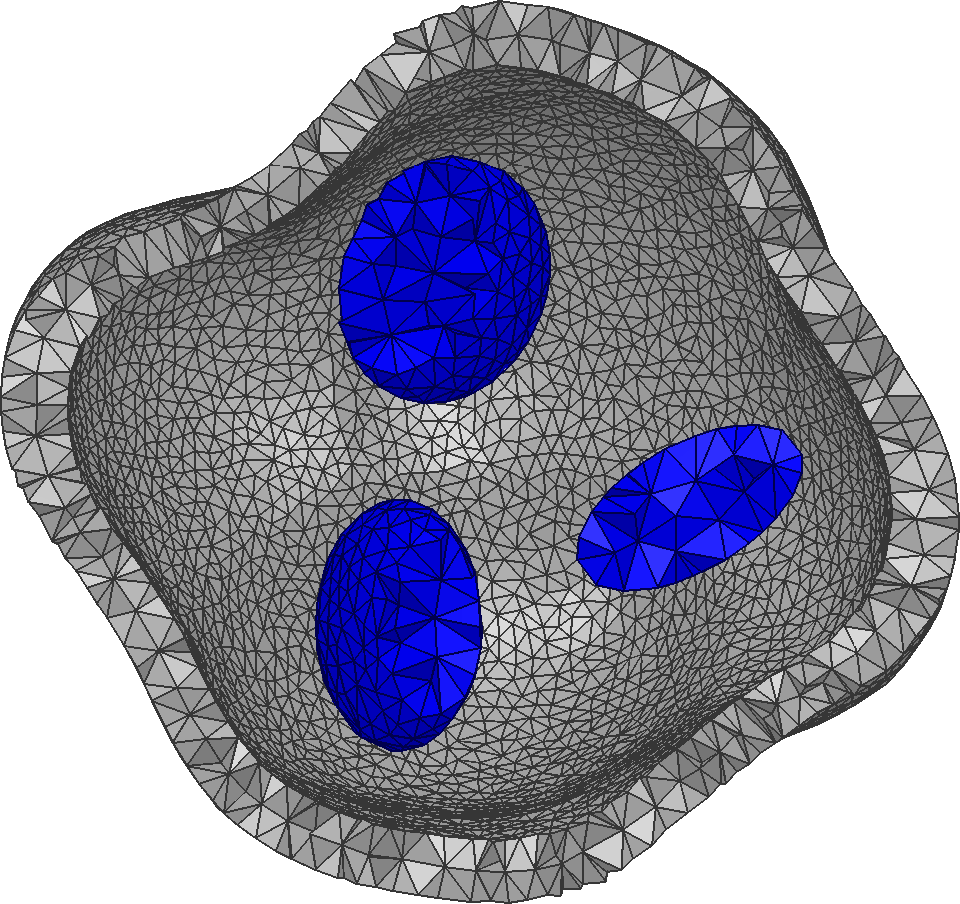}  \\ Set $\mathbf{S}$: Exact  \end{center} 
\end{minipage}
\begin{minipage}{3.5cm}\begin{center}
\includegraphics[width=3.0cm]{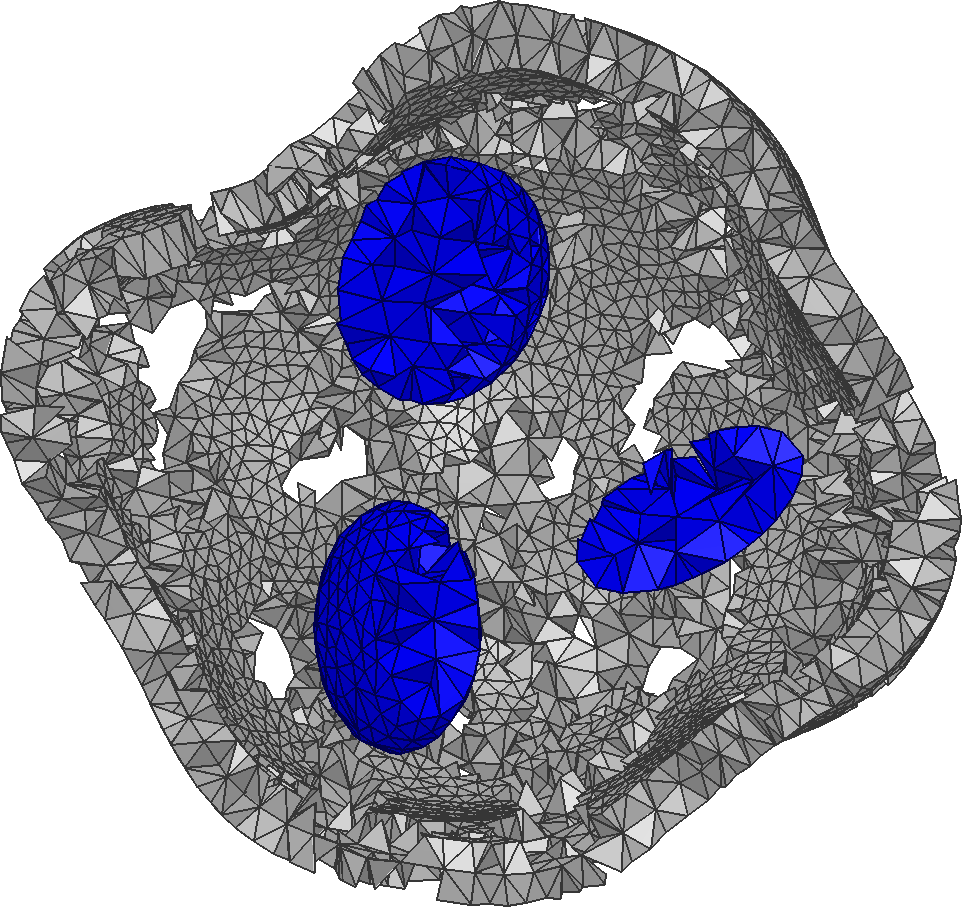}  \\  Set $\mathbf{V}$: {\bf (i)} 
 \end{center}
\end{minipage}
\begin{minipage}{3.5cm} \begin{center}  
\includegraphics[width=3.0cm]{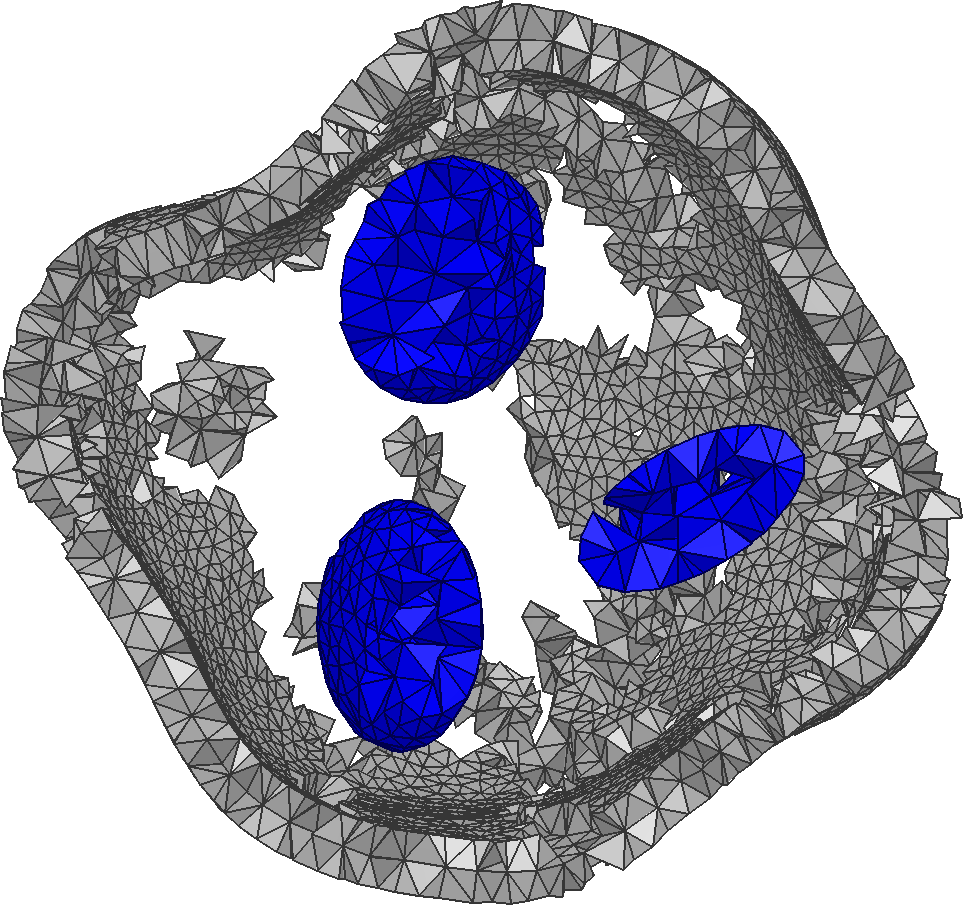}  \\  Set $\mathbf{V}$: {\bf (ii)}  \end{center} 
\end{minipage}
\begin{minipage}{3.5cm}\begin{center}
\includegraphics[width=3.0cm]{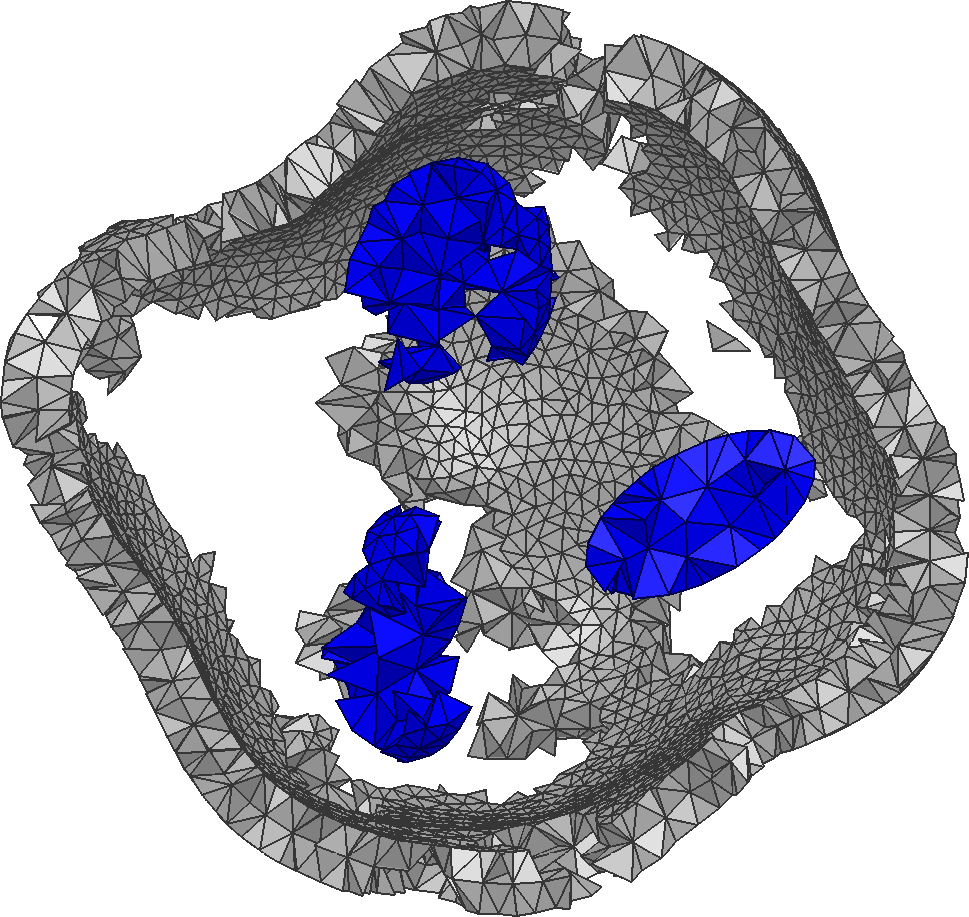}   \\  Set $\mathbf{V}$: {\bf (iii)}
 \end{center}
\end{minipage} 
\end{scriptsize} \\ \end{center} \end{framed} 
\begin{framed}
\begin{center} \begin{scriptsize}  
Clipping plane: $y = 0$ \\ \begin{minipage}{3.9cm} \begin{center}  
\includegraphics[width=3.0cm]{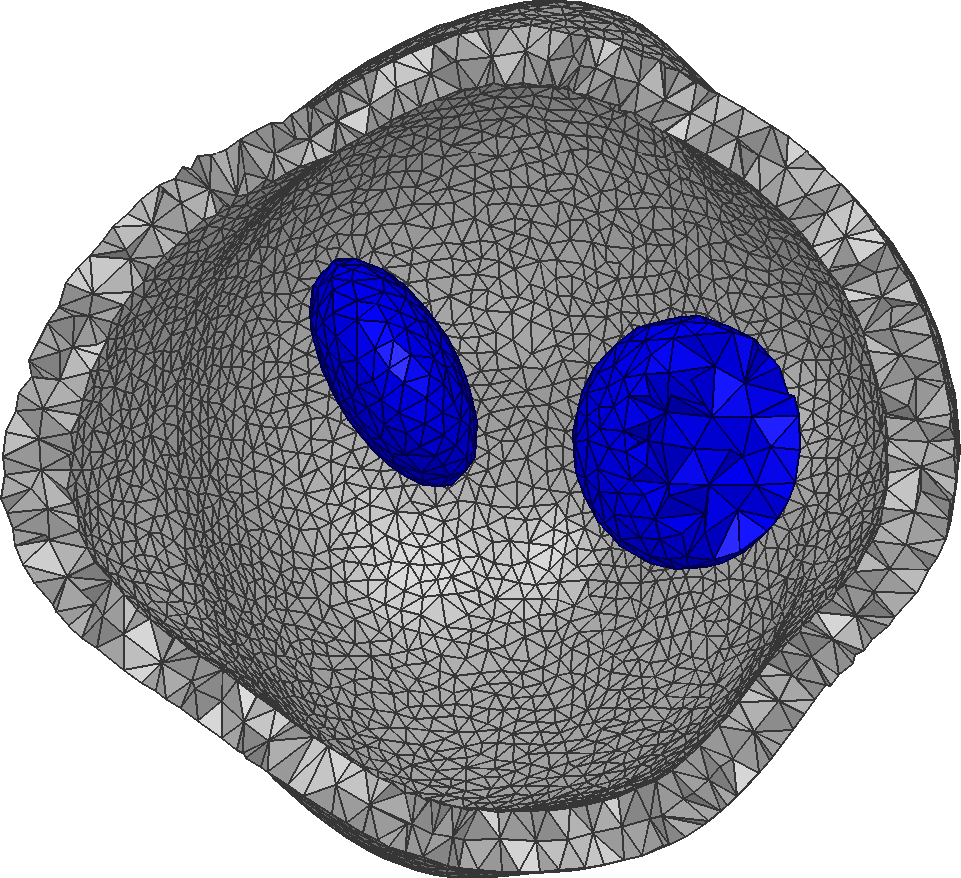}  \\ Set $\mathbf{S}$: Exact   \end{center} 
\end{minipage}
\begin{minipage}{3.5cm}\begin{center}
\includegraphics[width=3.0cm]{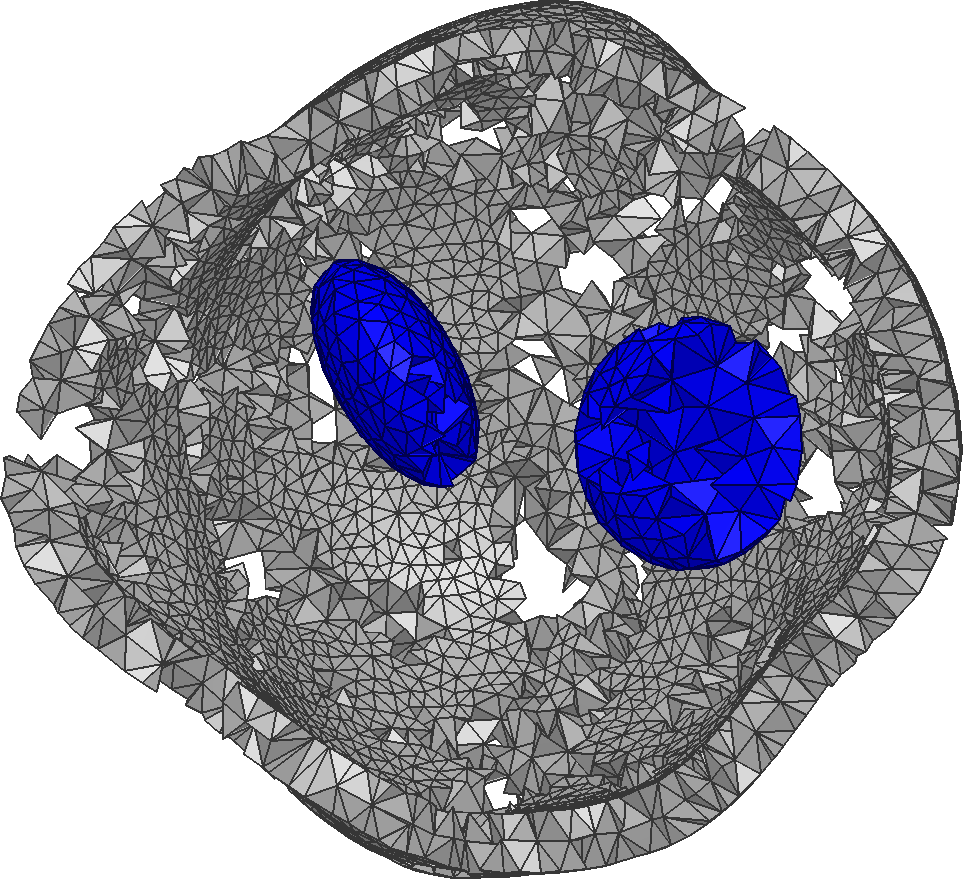}  \\  Set $\mathbf{V}$: {\bf (i)}
 \end{center}
\end{minipage}
\begin{minipage}{3.5cm} \begin{center}  
\includegraphics[width=3.0cm]{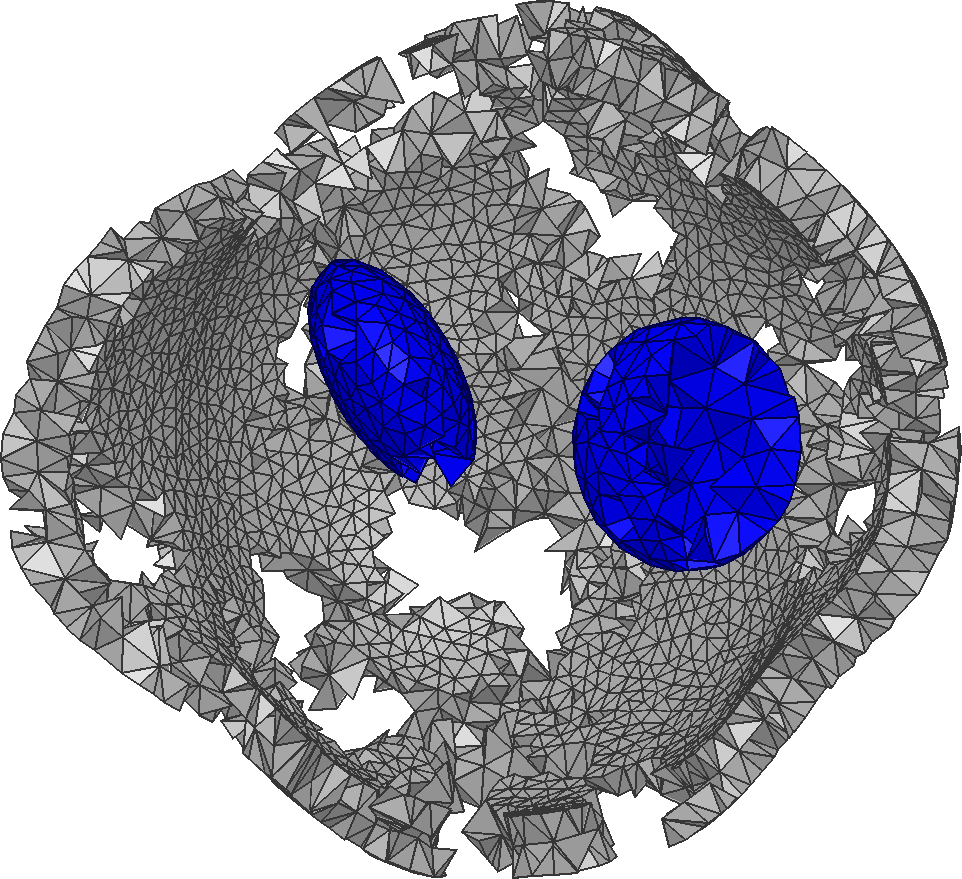}  \\ Set $\mathbf{V}$: {\bf (ii)}   \end{center} 
\end{minipage}
\begin{minipage}{3.5cm}\begin{center}
\includegraphics[width=3.0cm]{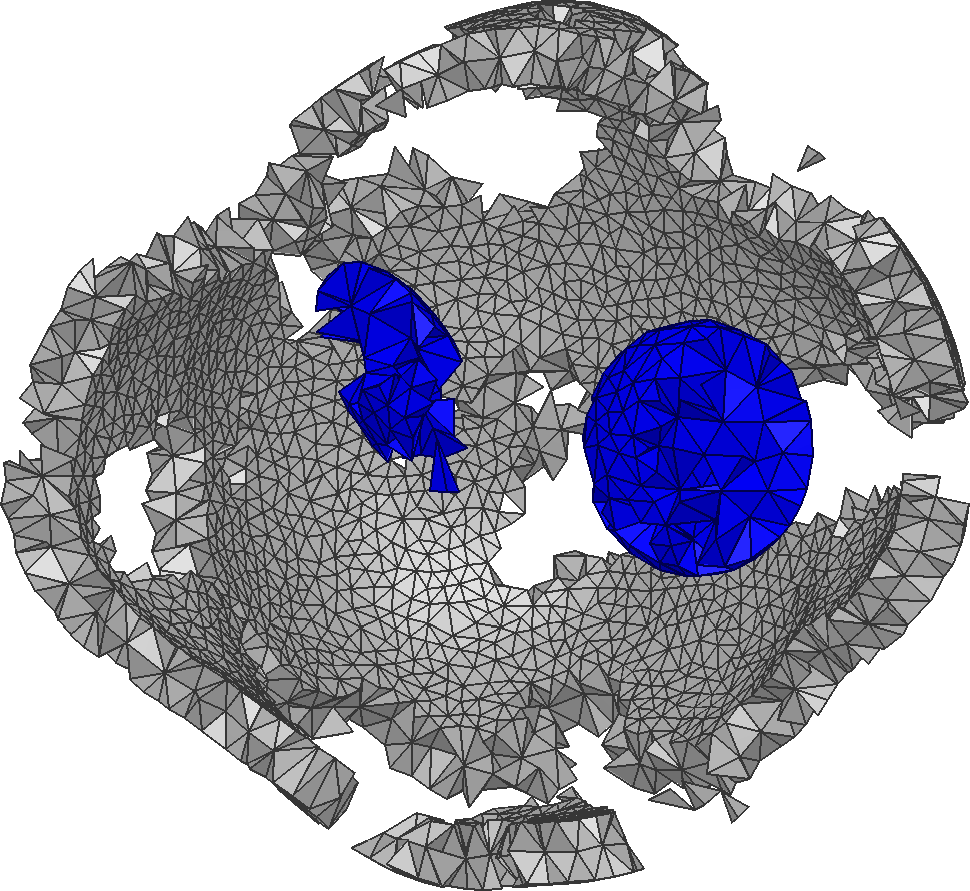}   \\ Set $\mathbf{V}$: {\bf (iii)}
\end{center}
\end{minipage}
\end{scriptsize} \\  \end{center} \end{framed} 
\begin{framed}
\begin{center} \begin{scriptsize} 
Clipping plane: $x = 0$ \\
\begin{minipage}{3.5cm} \begin{center}  
\includegraphics[width=3.0cm]{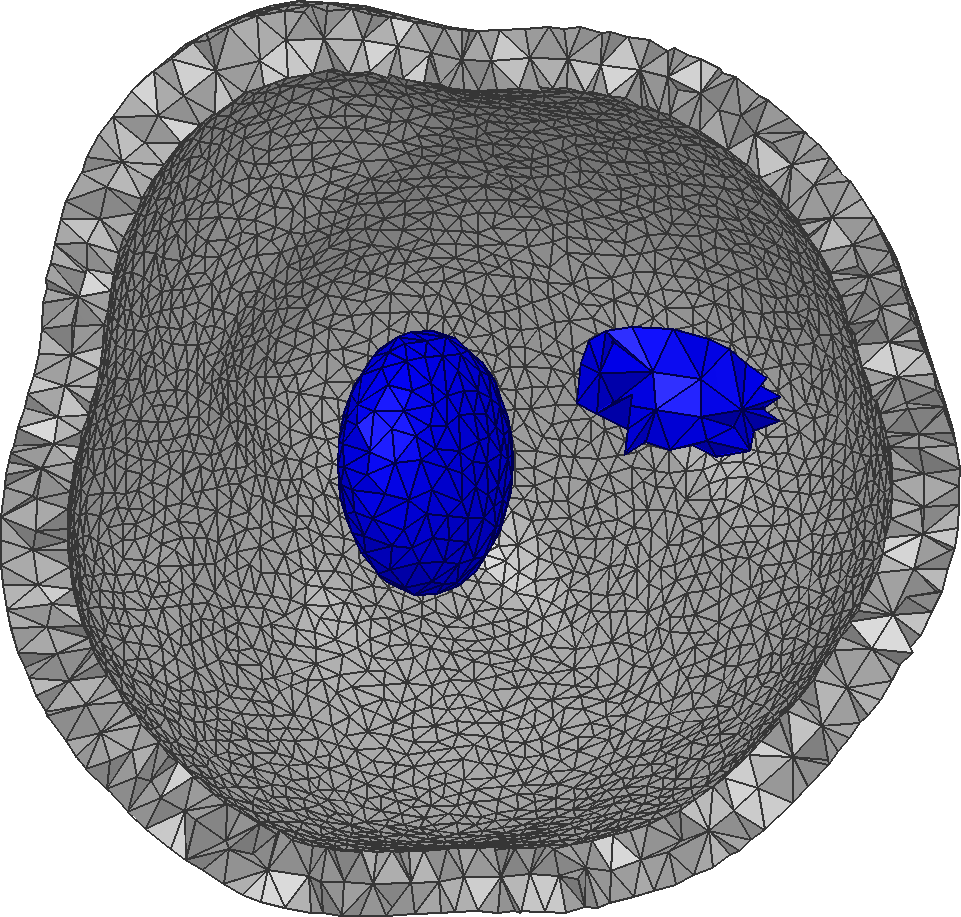}  \\ Set $\mathbf{S}$: Exact   \end{center} 
\end{minipage}
\begin{minipage}{3.5cm}\begin{center}
\includegraphics[width=3.0cm]{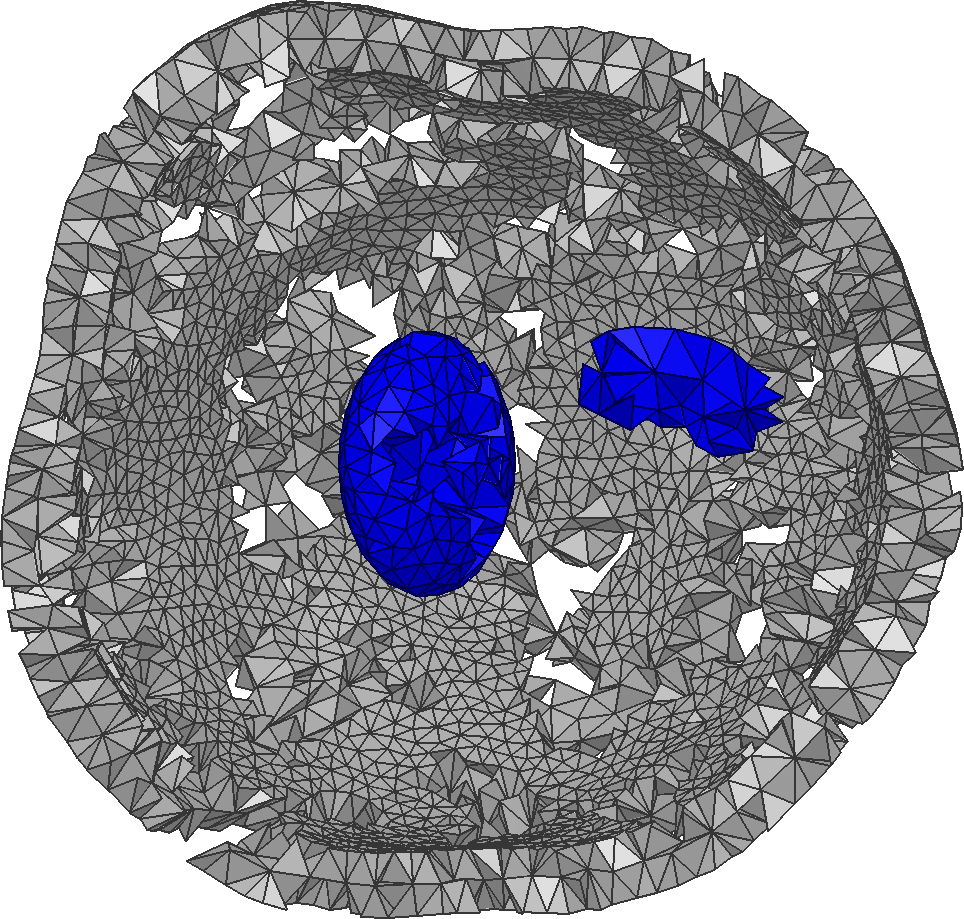}  \\ Set $\mathbf{V}$: {\bf (i)}
 \end{center}
\end{minipage}
\begin{minipage}{3.5cm} \begin{center}  
\includegraphics[width=3.0cm]{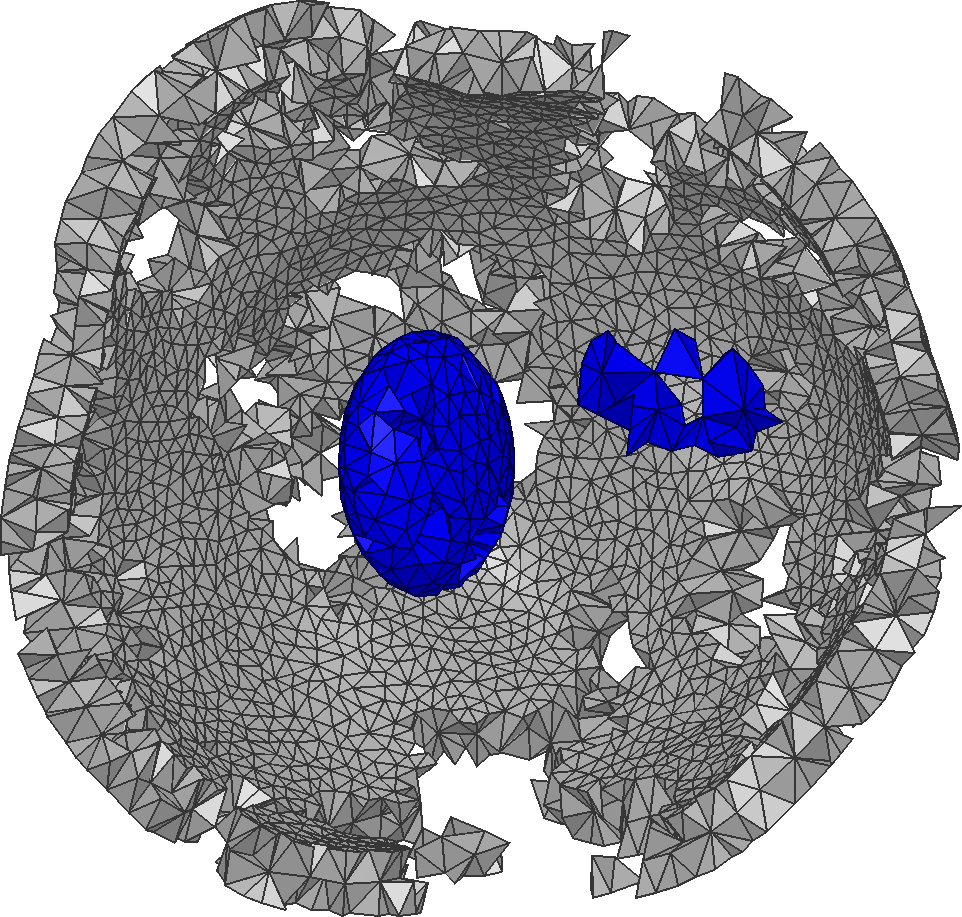}  \\ Set $\mathbf{V}$: {\bf (ii)}    \end{center} 
\end{minipage}
\begin{minipage}{3.5cm}\begin{center}
\includegraphics[width=3.0cm]{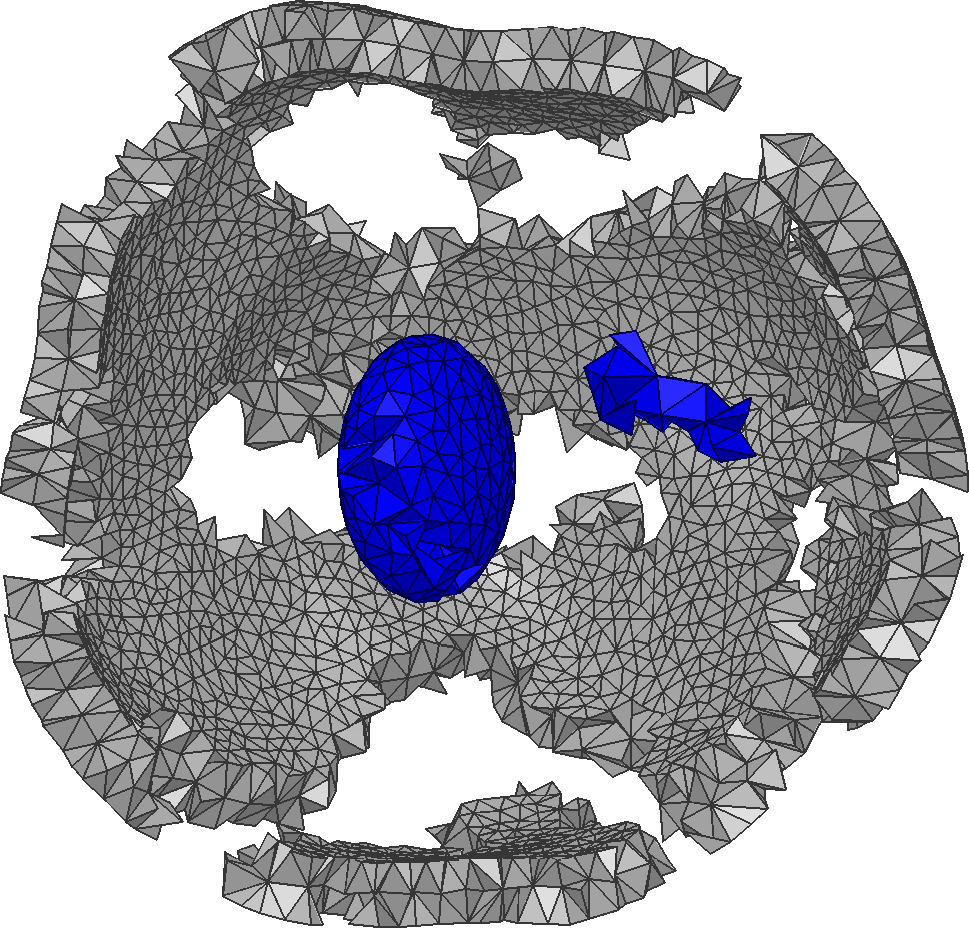}   \\ Set $\mathbf{V}$: {\bf (iii)}
 \end{center}
\end{minipage} 
\end{scriptsize}  
\end{center} \end{framed} 
 \end{framed} 
\caption{A visual comparison of the reconstructions {\bf (i)}, {\bf (ii)} and {\bf (iii)} obtained, respectively, with configurations (A), (B) and (C) using  bistatic data and total noise with the relative standard deviation of -15 dB. On each row, the left image illustrates  the  ROI  $\mathbf{S}$ (Section \ref{inversion_accuracy})  containing the exact surface dust layer (grey) and voids (blue). The other images present, the overlap set $\mathbf{V}$, i.e., the relative overlap between the exact permittivity distribution and the reconstruction (Section \ref{inversion_accuracy}).   
\label{result_comparison}}
\end{figure*}

\begin{table} \begin{center}
\caption{ROE and RVE for reconstructions {\bf (i)}, {\bf (ii)} and {\bf (iii)} obtained with bistatic data, total noise of -15 dB and configurations (A), (B) and (C), respectively (Figure \ref{result_comparison}). The values have been calculated separately for the total ROI $\mathbf{S}$ and for its intersection with the surface layer and voids. \label{ROE_RVE_table}}
\begin{tabular}{lllrrr} 
\hline
Rec.\ & Conf.\ & Measure & Total (\%) & Surface (\%) & Deep (\%) \\
\hline
{\bf (i)} & (A)     & ROE     & 33         & 34           & 21 \\
&        & RVE     & 22         & 5          & 48  \\
{\bf (ii)} & (B)     & ROE     & 39         & 41           & 20 \\
&        & RVE     & 34         & 27         & 57 \\        
{\bf (iii)} & (C)     & ROE     & 45         & 45          & 46 \\
    &    & RVE     & 62         &  55         & 85 \\ 
\hline
\end{tabular}
\end{center}
\end{table}

\section{Results}
\label{r}

The mathematical far-field model introduced in this paper was found to perform adequately for the test asteroid model with different levels of the total noise. The results have been included in Figures \ref{results_total}--\ref{result_comparison} and Tables \ref{ROE_RVE_table} and \ref{noise_table}. 

Figures  \ref{results_total}, \ref{results_surf} and \ref{results_deep} illustrate the ROE and RVE as a function of the total noise for the full ROI $\mathbf{S}$, the surface layer and voids, respectively.  At each investigated noise level, a sample of 30 different reconstructions, obtained with independent realizations of the total noise, has been visualized as a box plot bar. The overall error level can be observed to be elevated for the -5 and 0 dB total noise levels. In comparison between the monostatic and bistatic CRT, the latter was found to be more robust with respect to noise and limited-angle data. 

Figure \ref{result_comparison} visualizes reconstructions {\bf (i)}, {\bf (ii)} and {\bf (iii)} obtained with the total noise level -15 dB,  bistatic data and configurations  (A), (B) and (C), respectively. The errors of the limited-angle reconstructions {\bf (ii)} and {\bf (iii)}, are particularly concentrated around the z-axis, that is, around the aperture in the limited-angle measurement positions. In {\bf (iii)}, they are also notably more spread than in {\bf (ii)}. Table \ref{ROE_RVE_table} includes the ROE and RVE for {\bf (i)}, {\bf (ii)} and {\bf (iii)} calculated separately for the full ROI  $\mathbf{S}$ and for its intersection with the surface layer and the deep part (voids).

Table \ref{noise_table} includes the relative signal amplitudes noise levels for bistatic measurement. Both galactic noise and the upper estimate for the Solar radiation (active Sun) were  observed to stay at a tolerable level with median peak levels of -14 dB and -8 dB, respectively. The lower Solar radiation estimate (quiet Sun) was below the investigated noise range with the median -48 dB. 

\begin{table} \begin{center}
\caption{ The  minimum, maximum and median values (dB) of the relative signal amplitude measurement noise standard deviation  $\sigma_m$ with respect to the signal peak intensity estimated for bistatic data in the set of the simulated signals.  \label{noise_table}}
\begin{tabular}{llrrrr}
 &  Source & Min & Max & Median  \\
\hline 
Relative signal amplitude &  &  -136 & -120 & -130 \\ 
\hline 
Relative noise $\sigma_m$  & Galactic & -24 & -7 & -14 \\ 
&Active Sun & -18 & -1 & -8 \\ 
&Quiet Sun &  -58  & -41 & -48 \\
\hline 
\end{tabular} \end{center} 
\end{table}

\section{Discussion}
\label{d}

This paper introduced and validated a mathematical far-field model applicable in the Computed Radar Tomography (CRT) imaging \cite{persico2014, devaney2012} of small solar system bodies (SSSBs). In particular, the Deep Interior Scanning CUbeSat (DISCUS) mission concept \cite{deller2017,bambach2017} was examined as a potential application of this method. The numerical results obtained with the test asteroid model and the relative total noise range of -25--0 dB suggest that a sparse set of full-wave measurements can be inverted with the planned DISCUS mission  specifications. A sufficient reconstruction accuracy was obtained with sparse full- and limited-angle data.  Furthermore, the results suggest that a bistatic (dual spacecraft) measurement technique \cite{willis2008} with a fixed $25$ degrees angle between the transmitter and receiver can improve the reliability of the inversion as compared to the monostatic (single spacecraft) approach. 

In this study, the simulations were conducted for orbit radius of $5$ km. The final orbit is determined by the spacecraft's $\Delta v$ budget and orbit stability and is currently under investigation. Generally, the closer the approach the better is the signal-to-noise ratio of the measurements. We currently expect that a  radius of about 5 km will be achievable by a CubeSat. For comparison, Rosetta orbited  at around 10 km distance with its closest approach at about 4.5 km \cite{kofman2015} but was limited mostly by the environment of the active comet 67P/Churyumov-Gerasimenko. The present results suggest that a deeper descent is not needed for detecting a surface dust layer and deep interior voids. It is noteworthy that performing close observations of SSSBs is a recent tendency in the planetary research. For example, during Rosetta's final descent its Osiris wide-angle camera took its final image at 20 m altitude \cite{barbieri2017,clery2016}. Another example is the ongoing Osiris-REx mission \cite{berry2013, lauretta2012} in which  the goal is to achieve a 730 m orbiting distance and also to bring a sample of the regolith back to Earth.  Moreover, advanced active control strategies for  hovering in the vicinity of an SSSB have been developed  \cite{broschart2005, lee2016}. 

The present results suggest that an appropriate reconstruction quality can be achieved, if the standard deviation of the total noise is below -10 dB. Of the investigated sources of measurement errors \cite{barron1985, kraus1967}, the Sun's radiation during its active phase seems to exceed this limit for the 5 km orbiting distance which will need to be taken into account in the mission design. A natural way to reduce the noise would be to point the radar antennas towards the Sun. Namely, a dipole antenna is practically insensitive to radiation propagating along its axis. The galactic background noise, which cannot be reduced, seems to remain on an acceptable level with a median of  around -14 dB. Other potential noise sources not investigated in this study include for example Jupiter's radiation the magnitude of which depends largely on the target asteroid's position in the solar system. Furthermore, it seems obvious that achieving a feasibly low measurement noise with the 10 W transmitting power applied in this study will, in practice, require applying the stepped-frequency technique \cite{iizuka1984,gill2001,paulose1994} which allows  dividing the total radar bandwidth and, thereby, also the  noise, into narrow frequency lines. 
 
The errors related to the forward modeling can be significant and require further research. In the CONSERT measurements, the unpredicted noise peaks were observed to stay mainly -20 dB below the actual signal peak, suggesting that also those errors remain tolerable. Akin to \cite{pursiainen2014}, the  echo reflecting from the surface opposite to the spacecraft was found to be noisy. Achieving the best possible reconstruction quality necessitated excluding this echo from the measurement data. That is, the recorded time interval had to be limited to  5.0 $\mu$s. 
 
Based on the comparison between the full- and limited-angle tomography results, it seems that the interior structure can be reconstructed, if targeted NEA has a typical spin  without the need to alter the orbiting plane of the spacecraft which greatly simplifies operations. Namely, the spin latitude is close to -90 degrees for a large majority of the small NEAs \cite{la2004},  suggesting that a better measurement coverage than in the 30 degrees limited-angle test can be achieved. 

The bistatic CRT was found to an essential way to improve the inversion reliability as compared to the monostatic approach. The advantage of the bistatic measurement was observed to be particularly emphasized for a  high total noise level and sparse limited-angle observation both of which are potential scenarios for a space mission. Hence, we propose that  maximizing the reliability of the data requires measurements between two spacecraft in addition to recording the backscattering data at the point of transmission. 

In this study, the smallest (low-noise) ROE and RVE were in some cases obtained with the monostatic approach. This was obviously due to the larger polarization shift in the signal captured by the second spacecraft, following from the non-direct reflection. Consequently, spacecraft positioning can have a major effect on the signal quality. The present choice for the distance between the transmitter and receiver is based on our preliminary numerical tests using the test asteroid model. Further optimization can be done in the future. 

The current computational implementation is scalable and allows using any asteroid geometry. In a three-dimensional spatial scaling, the system matrix size is roughly proportional to $(s_1 / s_0)^3$, where $s_0$ and $s_1$ denote the scaling factors of the original and scaled domain. With the current measurement setting, the system size would be  approximately 64 GB, for the diameter of 1100 m, i.e., two times the current one. The 0.62 GB mass matrix of the current system would be of the size 10 GB for the scaled one. A system size of 64 GB is feasible regarding an implementation in a computation cluster. If the cluster nodes are equipped with a GPU with more than 10 GB memory, then the mass matrix can be inverted rapidly in the GPU which, due to its parallel computing capabilities, may be assumed to provide a faster solution for sparse matrix--vector multiplication than the central processing unit.

Based on the diameter distribution of the NEAs  \cite{trilling2017,mainzer2011}, the range of potential target diameters for DISCUS extends to at least 1000 m above which the distribution decays. For a   target diameter significantly larger than 1000 m, it might be reasonable to limit the imaging to some {\em a priori} estimated depth estimated, e.g., based on maximal observed signal penetration. If necessary, it is also possible to compress the memory consumption by replacing some of the matrices with matrix-free functions returning a given matrix--vector product. Also methodological development can be considered. For example, the discontinuous Galerkin time-domain method can be compared to the current finite element time-domain implementation  \cite{angulo2014,lu2005} and the leap-frog iteration can be replaced with the well-known Runge-Kutta algorithm. 

The reconstruction of interior structures containing less contrast than the structure analyzed in this study might also prove more difficult, and the existence of multiple scattering surfaces in the interior might increase the signal to noise required for a good reconstruction. It is therefore planned to assess the reconstruction of more complex interior structures, as for example an interior filled with spherical monolithic fragments following a power law size distribution as motivated in \cite{deller2015,deller2016}.

An ongoing future work is to verify the current results for a carefully simulated orbit. 
Additionally, the effect of the polarization and sparse limited-angle data on the reconstruction quality will be studied further. 
An  asteroid flyby can be considered as an alternative way to do tomographic measurements. Based on this study, achieving a sufficient signal-to-noise ratio might be challenging for a flyby, since flybys are usually made in the range of 1000 km. For example, the closest point of Rosetta's flyby at the asteroids Lutetia and Steins was 3170 km and 800 km \cite{accomazzo2012,keller2010}, respectively. On the contrary, CRT seems to require an extremely close 5 km rendezvous. Therefore, to achieve a reasonable signal to noise level during a flyby mission, concepts of very close flyby configurations with low relative velocity have to be developed.

\section*{Acknowledgements}
MT, MK and SP were supported by the AoF Centre of Excellence in Inverse Problems. MT and SP were funded by the Academy of Finland Key Project number 305055. Special thanks to Juha Herrala and Kari Suomela for support in computing resources. 

\appendix
\label{appendix_1}

The present wave equation and its weak form  can be derived from the (unitless) Maxwell's equations  {\setlength\arraycolsep{2pt} \begin{eqnarray} 
\nabla \cdot \varepsilon_r \vec{E}  & = & 0 \label{mx1} \\
\nabla \cdot \vec{B} & = &  0 \label{mx2} \\
\nabla \times \vec{E}  & = & -  \frac{\partial \vec{B}}{\partial t } \label{mx3} \\
\nabla \times \vec{B} & = & \vec{J} + \varepsilon_r \frac{\partial \vec{E}}{\partial t}   \label{mx4}
\end{eqnarray}} in which $\vec{E}$ and $\vec{B}$ denote the electric and magnetic field, respectively,  and $\vec{J} = \sigma \vec{E} + \vec{f}$ is the total current density with $\vec{f}$ denoting the current density of the antenna. The curl of the third equation (\ref{mx3}) can be written as 
\begin{equation}
\nabla \times \nabla \times \vec{E} = \nabla (\nabla \cdot \vec{E}) -  \nabla^2 \vec{E} = -  \sigma \frac{\partial \vec{E}}{\partial t} -  \varepsilon_r \frac{\partial^2 \vec{E}}{\partial t^2} - \frac{\partial \vec{f}}{\partial t}. \label{mx_spec}
\end{equation}
Expressing the electric field and the position vector in the component-wise form, i.e.,  $\vec{E} = (E_1, E_2, E_3)$ and $\vec{x}= (x_1, x_2, x_3)$, respectively, 
one obtains the following general form of the present wave propagation model:
\begin{equation}
 \varepsilon_r \frac{\partial^2 {E}_i }{\partial t^2} + \sigma \frac{\partial E_i}{\partial t}  -  \sum_{j = 1}^3 \frac{\partial^2 E_i}{\partial x_j^2} + \frac{\partial}{\partial x_i} \sum_{j=1}^3 \frac{\partial E_j}{\partial x_j} = - \frac{\partial f_i}{\partial t}.   \label{wave1}
\end{equation}
The first-order formulation of this equation is given by 
{\setlength\arraycolsep{2pt}\begin{eqnarray}\label{first-order_system}  \label{first_order_form} 
\frac{\partial {g}^{\,{(i)}}_{j}}{\partial t}  -   \frac{\partial E_i}{\partial x_j}  & = &   0,  \\ 
\varepsilon_r \frac{\partial {E}_i}{\partial t} + \sigma {{E}_i} - \sum_{j = 1}^3 \frac{\partial {g}^{\,(i)}_j}{\partial x_j} + \frac{\partial}{\partial x_i} \sum_{j=1}^3{g}^{\,(j)}_{j}  & =&  -   f_i,  
 \end{eqnarray} } in which the first equation holds for $i = 1,2,3$ and the second one for $i,j = 1, 2,3$. The entries of the vectors $\vec{g}^{\, (1)}, \vec{g}^{\, (2)}$ and $\vec{g}^{\, (3)}$ are given by ${g}^{\, (i)}_{j} = \int_0^t \frac{\partial E_i}{\partial x_j}(\tau, \vec{x}) \, d \tau$ for $j = 1, 2, 3$. Multiplying the first and the second equation of (\ref{first-order_system}) by the test functions  $\vec{w}^{(i)} \in  [L_2(\Omega)]^3$ and $v_i  \in H^1(\Omega)$, respectively, and integrating by parts yields the system 
{\setlength\arraycolsep{1 pt} \begin{eqnarray} \label{wf1} 0 & = &    
\frac{\partial }{\partial t} \sum_{j=1}^3  \int_{\Omega} \!\! {\vec g}^{\,(i)}_j  {w}^{\,(i)}_j \, \hbox{d} V \! \! \nonumber \\ & & -  \sum_{j=1}^3  \int_{\Omega} \!\! {w}^{\,(i)}_j  \frac{\partial E_i}{\partial x_j} \, \hbox{d}  V, \\  \label{wf2}
- \int_\Omega f_i v_i \, \hbox{d} V  & = &   \frac{\partial}{\partial t} \!  \int_{\Omega} \! \! \varepsilon_r  \, E_i \, v_i \, \hbox{d}  V \!\! + \!\! \int_{\Omega} \! \! \sigma  \, E_i \, v_i \, \hbox{d}  V \!\!  \nonumber\\ & & +  \sum_{j=1}^3  \int_{\Omega} \!\!  {g}^{\,(i)}_j  \frac{ \partial{v_i}}{\partial x_j}  \, \hbox{d}  V \nonumber \\ 
& & -  \!\! \sum_{j=1}^3 \int_{\Omega} \!\!  {g}^{\,(j)}_{j}   \frac{ \partial{v_i}}{\partial x_i} \, \hbox{d}  V.    
\end{eqnarray}}
This system can be discretized via the approach presented in the two-dimensional study \cite{pursiainen2016}. Using the notation of \cite{pursiainen2016}, the last right-hand side term affecting the polarization of the wave, absent in the 2D case, is of the form 
$- {{\bf B}^{(i)}}^T \sum_{j = 1}^3 {\bf q}_k^{(k)}$, where ${\bf q}_j^{(i)}$ denotes the coordinate vector of ${g}_i^{(j)}$.

\bibliographystyle{IEEEtran}
\bibliography{pursiainen,references}

\medskip
Received xxxx 20xx; revised xxxx 20xx.
\medskip

\end{document}